\documentclass[journal ]{new-aiaa}
\usepackage[utf8]{inputenc}
\usepackage{textcomp}

\usepackage{graphicx}
\usepackage{amsmath}
\usepackage[version=4]{mhchem}
\usepackage{siunitx}
\usepackage{longtable,tabularx}
\newlength{\equalssep} 

\usepackage{amsfonts}
\usepackage{mathtools}   

\usepackage{mathrsfs}
\usepackage{array}
\usepackage[caption=false,font=footnotesize,labelfont=sf,textfont=sf]{subfig}
\usepackage{textcomp}
\usepackage{stfloats}
\usepackage{url}
\usepackage{verbatim}
\usepackage{svg}
\usepackage{multirow}
\usepackage{bm}
\usepackage{graphicx}
 \usepackage{placeins}
\usepackage{stfloats}
\usepackage{tabularx}
\usepackage{array}
\usepackage{amsmath}
\usepackage{graphicx}
\usepackage{hyperref}

\usepackage{algorithm}
\usepackage{algpseudocode}

\graphicspath{{Images/}}

\title{Real-time Priority-aware UAS Arrival Planning Framework for Multi-layer UTM}

\author{%
Anisha~Mohapatra\textsuperscript{$\ast$,1},
Prasanna~Ramamurthy\textsuperscript{$\ast$,2}, ~
Shridhar~Velhal\textsuperscript{3},
~M.~Sethumadhavan\textsuperscript{2},
and ~Suresh~Sundaram\textsuperscript{1}\\
\vspace{8pt}

\textsuperscript{1}Department of Aerospace Engineering, Indian Institute of Science, Bengaluru, India.\\

\textsuperscript{2}TIFAC-CORE in Cyber Security, Amrita School of Engineering, Coimbatore, Amrita Vishwa Vidyapeetham, India.\\
\textsuperscript{3}Robotics and AI Group, Lule{\aa} University of Technology, Lule{\aa}, Sweden. \\
\textsuperscript{$\ast$} equal contribution\\
}

\begin{document}

\maketitle
\begin{abstract}
This paper proposes a real-time, priority-aware UAS arrival-planning framework for multi-layer UTM terminal airspace represented as a funnel of concentric rings, discrete holding slots, connecting edges, and a landing pad. Managing dense arrivals is challenging because terminal resources are limited and UAS must remain spatially and temporally separated throughout their approaches. New arrivals, wind disturbances, and emergency landing requests further require rapid replanning beyond conventional landing-pad sequencing. To achieve computational tractability, the framework employs a greedy planning method that first establishes a priority-aware landing order and then plans the UAS sequentially. For each UAS, the method determines its slot assignments, traversal path, landing time, and slot entry and exit times while treating previously planned trajectories as fixed. Spatial and temporal constraints prevent conflicts on holding slots, connecting edges, and the landing pad. Within a rolling-horizon implementation, the framework incorporates new arrivals, reactively adjusts plans affected by wind disturbances, and prioritizes emergency UAS without altering completed operations. Computational results show that the method generates and updates feasible, conflict-free arrival plans within operational time limits while providing expedited landing access to emergency UAS. The results demonstrate the suitability of the proposed framework for responsive and safety-critical arrival management in dynamic multi-layer UTM environments. 
Video link:  \href{https://youtu.be/F9IP1dz-9PE}{https://youtu.be/F9IP1dz-9PE}
\end{abstract}

\section*{Nomenclature}

{\renewcommand\arraystretch{1.0}
\settowidth{\equalssep}{\quad=\quad}

\noindent
 
\begin{longtable*}{@{}p{0.16\linewidth}@{\quad=\quad}p{\dimexpr0.80\linewidth-\equalssep\relax}@{}}
$\mathcal{U}$ & set of arriving UAS ; index $i$ \\   
$N \coloneqq |\mathcal{U}|$ & number of arriving UAS \\
$L \in \mathcal{L} $   & ring level index  $\mathcal{L}  = \{0,1,\cdots, L_{\max}\}$  \\  
 $J_L$ & set of angular slots on ring $R_L$ (e.g.,  $J_1=\{N,E,S,W\}$ for $L=1$;.)\\
$\mathrm{ETA}_i \in \mathbb{R}_{\ge 0}$ & earliest time UAS $i$ may enter the outermost ring \\    
$\rho_i\in\mathbb{Z}_{\geq 0}$ & greedy sequencing priority of UAS $i$, where a larger value indicates that the UAS is planned earlier \\
$r_L\in\mathbb{R}_{\ge0}$ & radius of ring level $R_L$ \\
$z_L\in\mathbb{R}_{\ge0}$ & altitude of level $L$\\
$\theta_{j,L}$ & angular position of slot $j\in J_L$ at level $L$\\ 
$t_i^{L} \in \mathbb{R}_{\ge 0}$ & start/entry time of UAS $i$ at ring $R_L$\\
$e_i^{L}\in\mathbb{R}_{\ge 0}$  & exit time of UAS $i$ from its selected slot on ring $R_L$ \\
$h_{i,L}\in\mathbb{R}_{\ge 0}$ & holding time of UAS $i$ at its selected slot on ring $R_L$ \\  
$h_{\min,L}\in\mathbb{R}_{\ge 0}$ & minimum holding time at level $L$\\  
$\delta^{(L)}_{j_L\to j'_{L-1}}\in\mathbb{R}_{\ge 0}$ & travel time from slot $j_L\in J_L$ on level $L$ to slot $j'_{L-1}\in J_{L-1}$ on level $L-1$ \\  
$s_0 := h_0 + \Delta^{(0)}$ & minimum separation between consecutive pad start times \\  
$f_{j_L,j'_{L-1}}\in\{0,1\}$ & angular feasibility  whether the transition from slot $j_L$ to slot $j'_{L-1}$ is within the prescribed $\pm\theta_{\max}$ funnel angle or not \\  
$\Delta^{(L)}\in\mathbb{R}_{\ge0}$ & safety headway between consecutive uses of any slot on $R_L$\\  
$x_{i,j,L} \in \{0,1\}$ & is the decision whether UAS $i$ is assigned to slot $j$ at level $L$ or not \\
$y^{(L)}_{i,j_L,j'_{L-1}}\in\{0,1\}$ & equals $1$ iff UAS $i$ traverses the directed descent edge from slot $j_L\in J_L$ to slot $j'_{L-1}\in J_{L-1}$\\ 
$o_{i\ell jL}\in\{0,1\}$ &binary decisoin variable whether   UAS $i$ uses slot $j$ in ring $R_L$ after UAS $\ell$ or not \\  
$\beta_{j_1}\ge 0$ & optional exogenous allowance into inner slot $j_1$ ($\beta_{j_1}{=}0$ for a closed funnel)\\ 
$\mathbf{V}_{w}\in\mathbb{R}^{3}$ & constant wind velocity vector \\  
$V_{w,z}$ & vertical component of the wind velocity\\
$v^{\mathrm{eff},(L)}_{j_Lj'_{L-1}}\in\mathbb{R}_{>0}$ &wind-adjusted effective ground speed along the descent path from slot $j_L$ to slot $j'_{L-1}$  
  \end{longtable*}}
  
\addtocounter{table}{-1}

\section{Introduction}
Urban Air Mobility (UAM) is envisioned as a key component of future intelligent transportation systems, enabling on-demand passenger and cargo transportation through electric vertical take-off and landing (eVTOL) operating in low-altitude urban airspace. Continued advances in electric propulsion, autonomous flight, and communication technologies are expected to substantially increase the density of eVTOL operations over urban areas.
As highlighted by recent demand forecasting and planning studies, the successful deployment of Advanced Air Mobility (AAM) depends not only on vehicle technologies but also on the efficient utilization of vertiports, terminal airspace, and supporting infrastructure under increasing traffic demand \cite{acharya2026demand}. Consequently, terminal arrival management is expected to become one of the primary operational bottlenecks in future high-density AAM systems.

To address these challenges, research has progressed along several complementary directions. Network-level scheduling and UTM studies have investigated fleet scheduling, conflict resolution, traffic management, and strategic airspace utilization to improve overall operational efficiency \cite{wei2021scheduling,rigas2024scheduling,tang2023incorporating,meng2025optimization,tan2019evolutionary}. Parallel efforts have focused on trajectory planning, energy-efficient flight, cooperative path planning, and contingency management for individual UASs and eVTOLs, improving flight efficiency, safety, and operational resilience under nominal and abnormal conditions \cite{pradeep2019energy,wang2026spatiotemporally,pinto2025towards,altun2023comprehensive,kim2026runtime,saez2025automated,jiang2026bi}. While these studies substantially improve large-scale airspace operations and individual vehicle performance, they primarily optimize operations before UAS enter the terminal arrival region and therefore provide limited support for coordinating multiple UAS competing for shared terminal-airspace resources immediately before landing.

Recognizing this limitation, increasing attention has recently shifted toward terminal operations. Existing studies have investigated vertiport infrastructure management, including landing-pad allocation, FATOs, gates, taxiways, digital twins, airspace reconfiguration, and vertiport capacity assessment \cite{chen2024integrated,saxena2026vertiport,baleghi2025real,zhang2025bigraph,vascik2020allocation,hu2024airspace,chang2026vertiport,li2025capacity}. These studies significantly improve infrastructure utilization, operational awareness, and terminal throughput. However, they primarily optimize either ground resources or evaluate airspace capacity, while UAS movement through the terminal arrival corridor itself remains largely predefined. Consequently, the airborne arrival process is generally treated as a transit phase rather than as a schedulable resource requiring coordinated spatial and temporal allocation.

The studies most closely related to the present work address terminal arrival organization and approach sequencing. Song et al. \cite{song2021approach} introduced the Vertiport Terminal Control Area (VTCA), proposing concentric holding circles and holding points together with the Sequence-Based Approach (SBA) and Branch-Queuing Approach (BQA) to organize eVTOL arrivals. This work established the fundamental concept of structured terminal airspace by exploiting the hovering capability of eVTOLs instead of conventional runway holding procedures. Building upon this concept, heuristic, mixed-integer linear programming, and rolling-horizon scheduling methods were developed to optimize eVTOL arrival sequences while satisfying separation and operational constraints \cite{pradeep2018heuristic,kleinbekman2018evtol,kleinbekman2020rolling}. More recently, Wei et al. \cite{wei2025optimization} extended ring-based approach management by introducing AIR, SRMA, and ARAM procedures capable of handling flight-priority differences and traffic-flow imbalance through multi-objective optimization. 
These studies clearly demonstrate the effectiveness of structured terminal airspace and priority-aware sequencing in improving arrival efficiency. Nevertheless, they continue to formulate terminal arrivals primarily as approach-procedure and sequencing problems. UAS progression is governed by predefined movement rules, while the terminal arrival airspace itself is not formulated as a unified optimization problem. Moreover  wind disturbance induces the variations in travel times, and relative velocity of UAVs; these effects are not explicitly incorporated. Consequently, coordinated allocation of holding slots across multiple altitude levels, conflict-free utilization of shared descent corridors, integrated scheduling of airborne and landing resources,  wind-aware adjustment of arrival trajectories and slot occupancy times, and system-wide redistribution of delay to accommodate high-priority UAS remain largely unaddressed. 

Despite significant advances in network scheduling, trajectory planning, terminal resource management, and arrival sequencing, existing studies optimize these components largely in isolation. In particular, even the closest ring-based arrival scheduling methods do not treat the terminal arrival corridor as a collection of shared spatiotemporal resources whose occupancy, transitions, and access must be jointly coordinated. As traffic density increases, multiple eVTOLs simultaneously compete for limited holding locations, descent corridors, and landing infrastructure, making the terminal arrival airspace itself a capacity-constrained system. Existing formulations therefore cannot simultaneously optimize spatial slot allocation, temporal progression through successive holding levels, exclusive utilization of shared descent corridors, landing-pad access, and priority-aware delay redistribution within a single optimization framework.

To address these limitations, this paper proposes a multi-altitude funnel-based spatiotemporal scheduling framework that models the terminal arrival airspace as a hierarchy of interconnected shared resources. Exploiting the omnidirectional landing capability of eVTOLs, a mixed-integer linear programming formulation jointly determines ring-slot assignments, holding durations, inter-ring transitions, and landing sequences while enforcing slot-capacity constraints, shared descent-corridor exclusivity, spatially feasible transitions, and wind-adjusted transit-time constraints.
The framework explicitly incorporates emergency-landing priorities, allowing high-priority UAS to be assigned earlier landing opportunities through coordinated reallocation of ring slots, descent corridors, and landing times. The resulting delay is redistributed across lower-priority UAS while maintaining spatial and temporal separation throughout the funnel.
Rather than merely resequencing UAS, the proposed framework coordinates the complete arrival process by jointly scheduling airborne and landing resources. Thus, the proposed framework facilitates emergency landings under wind disturbances while maintaining efficient terminal airspace utilization in dense traffic conditions.

The primary contributions of this paper are summarized as follows.
\begin{itemize}
 \item Unified spatiotemporal scheduling framework: A mixed-integer linear programming formulation is developed to jointly optimize spatial slot allocation, temporal progression, descent transitions, and landing-pad scheduling within a multi-altitude funnel architecture.
 \item Conflict-aware terminal-airspace coordination: The proposed formulation introduces integrated constraints for ring-slot capacity, shared descent-corridor exclusivity, spatial feasibility, and wind-adjusted transit-time parameterization, enabling conflict-free utilization of terminal-airspace resources.
 \item Priority-aware arrival management: A priority-weighted scheduling framework is proposed in which urgent UASs receive preferential landing access through coordinated redistribution of delays across lower-priority traffic rather than by simple resequencing.   
\end{itemize}

\section{Problem Formulation}
\begin{figure}[tp]
    \centering
    \includegraphics[width=\linewidth]{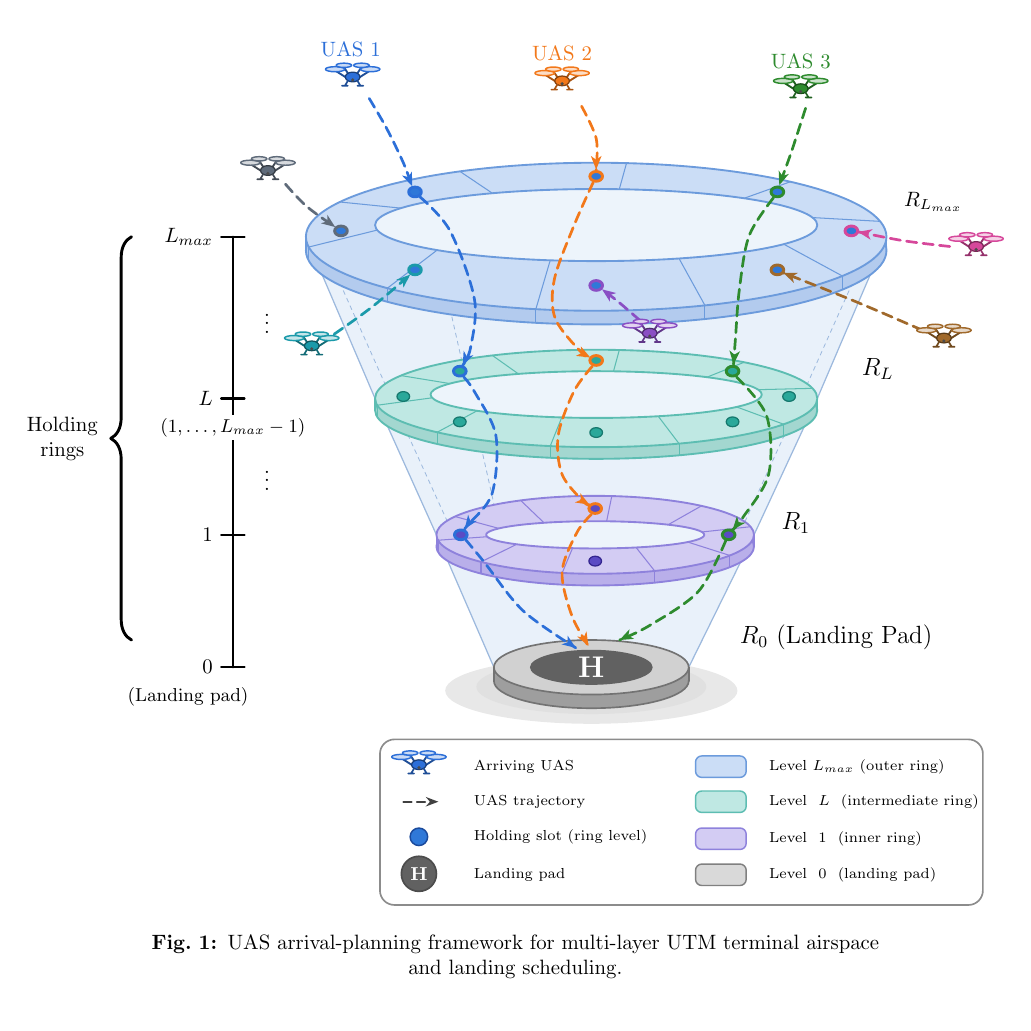}
    \caption{Multi-layer UTM airspace sectors and funnel-type structure for landing   }
    \label{fig:concept_funnel}
\end{figure}

We consider the scheduling of arriving UASs in a multi-level vertiport 
terminal area, as illustrated in Fig.~\ref{fig:concept_funnel}. Let $\mathcal{U}$ denote the set of arriving UAS, with $N=|\mathcal{U}|$, and let $L_{\max}$ denote the number of holding rings in the funnel. The terminal airspace is represented by the set of operational levels $  \mathcal{L} = \{L_{\max},L_{\max}-1,\ldots,1,0\}, $  where $R_{L_{\max}}$ is the outermost holding ring, $R_L$,  $L\in\{1,\ldots,L_{\max}-1\}$, represents an intermediate holding ring,  and $R_0$ represents the landing pad. Thus, the funnel contains $L_{\max}$ holding rings and one landing level.

Each holding ring $R_L$, $L\in\mathcal{L}\setminus\{0\}$, contains a finite set of admissible angular positions, or slots, denoted by $J_L$,  while the landing pad is represented by the singleton set $J_0=\{Pad\}$.  An arriving UAS $i\in U$ may enter the terminal airspace at the  outermost ring $R_{L_{\max}}$ no earlier than its earliest admissible  arrival time, $\mathrm{ETA}_i$. It then progresses monotonically toward the landing pad through the ordered sequence of rings.

At each visited holding ring, the UAS is assigned exactly one slot, may remain at that slot for a finite holding time, and subsequently traverses a feasible connection to a slot at the next lower level. The transit time from slot $j\in J_L$ to slot $k\in J_{L-1}$ is denoted by $\delta_{jk}^{L}$, for all $L\in\mathcal{L}\setminus\{0\}$, and is precomputed for every admissible slot pair. This transit time depends on the corresponding path geometry, the prescribed UAS speed, and the prevailing wind conditions.

The decision problem is therefore to determine, for each UAS, the selected slot at every level, the corresponding entry and exit times, the holding duration on each ring, and the relative service order whenever two UAS use the same slot or the landing pad. The binary variable \(x_{i,j,L}\) indicates whether UAS \(i\) is assigned to slot \(j \in J_L\) at level \(L\). The continuous variables \(t_i^{L}\), \(e_i^{L}\), and \(h_{i,L}\) denote the entry time, exit time, and holding time of UAS \(i\) at level \(L\), respectively. Additional binary ordering variables are introduced to enforce non-overlap and minimum separation on shared resources.

The formulation must satisfy both spatial and temporal feasibility. Spatially, each UAS can occupy only one slot at a given level, and transitions from the outer ring to the inner ring are restricted by the prescribed funnel geometry. Temporally, the exit time at any ring level is given by the sum of entry time and holding time, while the entry time at the next lower level is obtained by adding the corresponding inter-level transit time. Capacity constraints are imposed on every ring slot so that no two UAS overlap in time at the same slot, and a minimum headway is maintained between consecutive uses. The landing pad is modeled as a single-server resource with mandatory separation between successive touchdown operations.

The objective is to minimize the average priority-weighted delay from the earliest admissible arrival time of each UAS to its touchdown time on the landing pad. In this way, the formulation jointly captures spatial assignment, temporal sequencing, holding, and separation requirements within a unified scheduling framework. The resulting model is a mixed-integer linear optimization problem for conflict-free and operationally feasible vertiport arrival scheduling.

\section{Methodology}



\subsection{Objective function}
The objective minimizes the average weighted delay from each UAS's arrival time at the terminal area to its touchdown time on the landing pad. This promotes earlier and more efficient landing schedules while accounting for UAS priority. Mathematically,
\begin{align}
\min J=\frac{1}{N}\sum_{i\in U}\alpha_i\left(t_i^{0}-\mathrm{ETA}_i\right) \label{eq:objective}
\end{align}

\subsection{Constraints}
\subsubsection{Unique position in a ring}
A UAS can occupy only one position at each ring level. This enforces the UAS to move towards the landing pad and avoids the shuffling of positions within a ring. Mathematically,
\begin{subequations}\label{eq:unique_position_ring}
\begin{align}
\sum_{j \in J_{L}} x_{i,j,L} &= 1, 
\quad \forall L \in \mathcal{L}
\label{eq:unique_position_ring_a}\\
t_i^{L_{\max}} &\ge \mathrm{ETA}_i
\label{eq:unique_position_ring_b}
\end{align}
\end{subequations}


\subsubsection{Continuity through the rings}
This constraint links the entry, exit, and descent times across successive ring levels. It ensures that any delay is absorbed as holding at the assigned slot, while the travel time between two consecutive levels remains fixed. 

The end time of UAS $i$, is sum of start time and the hold time in that layer, mathematically.
\begin{align}
t_i^{L-1}
&=
e_i^{L}
+
\sum_{\substack{j_L \in J_L\\ j'_{L-1}\in J_{L-1}}}
\delta^{(L)}_{j_L \rightarrow j'_{L-1}}
y^{(L)}_{i,j_L,j'_{L-1}},
\nonumber\\
&\hspace{2.5em}
\forall i \in U,\;
\forall L \in \mathcal{L}\setminus\{0\}.
\label{eq:continuity_through_rings_c1}
\end{align}

The entry time at the next lower level is the exit time from the current level plus the transit time along the selected descent path. Since the transit time depends on the chosen origin and destination slots, the continuity relation is written as
\begin{align}
t_i^{L-1}
&=
e_i^{L}
+
\sum_{j_L \in J_L}
\sum_{j'_{L-1} \in J_{L-1}}
\delta^{(L)}_{j_L \rightarrow j'_{L-1}}
y^{(L)}_{i,j_L,j'_{L-1}}, \label{eq:continuity_through_rings_c} \\
&\quad
\forall i \in U,\; \forall L \in \mathcal{L}\setminus\{0\} \nonumber
\end{align}

\subsubsection{Ring--slot capacity}
A slot can be assigned to at most one UAS at a time, and a minimum headway is maintained between two consecutive uses of the same slot. The binary ordering variable determines which UAS is scheduled first when both select the same slot.

A UAS $i$, is considered in the capacity of a slot only if it is assigned to the slot. 
 \begin{align}
& o_{i\ell j L} \le x_{i,j,L}   , \quad   o_{i\ell j L} \le x_{\ell,j,L} 
\label{eq:ring_slot_capacity_d} \\
&\quad \forall i,\ell \in U,\; i\neq \ell,\; \forall L \in \mathcal{L},\; \forall j \in J_L \notag 
\end{align}
 
For  UAS $i$ and $\ell$, only one UAS can be visit slot $j$ before another,  mathematically, 
 \begin{align}
& o_{i\ell jL} + o_{j \ell i L} \le 1
\label{eq:ring_slot_capacity_e}\\
& \quad \forall i,\ell \in U,\; i\neq \ell,\; \forall L \in \mathcal{L},\; \forall j \in J_L \notag 
\end{align}
 
For UAS $i$ and $\ell$, and slot $j$ in layer $L$, if both UAS visits the slot $j$ then 
 \begin{align}
& x_{i,j,L} + x_{\ell,j,L} - 1 \le o_{i\ell jL} + o_{\ell ijL}
\label{eq:ring_slot_capacity_f}\\
&\quad \forall i,\ell \in U,\; i\neq \ell,\; \forall L \in \mathcal{L},\; \forall j \in J_L \notag
\end{align}
 
If UAS $i$ and $\ell$, both visits slot $j$ in layer $L$,  then the sequence of these two UASs are constrained using big M method. The start time of later UAS is more than the end time of previous UAS and safety headway of that layer, mathematically, 
\begin{align}
&t_i^{L}  \ge t_{\ell}^{L} + h_{\ell,L} + \Delta^{(L)} - M\left(1-o_{i\ell jL}\right)
\label{eq:ring_slot_capacity_a}\\
& t_{\ell}^{L}  \ge t_i^{L} + h_{i,L} + \Delta^{(L)} - M\,o_{i\ell jL}
\label{eq:ring_slot_capacity_b}\\
&\quad \forall i,\ell \in U,\; i\neq \ell,\; \forall L \in \mathcal{L},\; \forall j \in J_L \notag
\end{align}

\subsubsection{Incoming-edge exclusivity}
Each ring slot may be reached through multiple incoming descent edges from the immediately higher level. To avoid conflicts during descent, only one incoming edge leading to a given destination slot may be active at any instant. Since an UAS can begin traversing an incoming edge only after departing its assigned slot on the upper level, the upper-level exit time is treated as the edge-entry time. Rather than introducing additional edge-ordering variables, the existing slot-ordering variable $o_{i\ell jL}$ is reused. Consequently, if UAS $U_i$ is scheduled before UAS $U_\ell$ at destination slot $j$, then UAS ($U_\ell$) cannot begin traversing any incoming edge toward that slot until UAS $U_i$ has completely vacated the destination slot.

Using the big-(M) formulation, the incoming-edge exclusivity constraints are written as
\begin{align}
e_{\ell}^{L+1} &\ge e_i^{L} - M\left(1-o_{i\ell jL}\right), \label{eq:edge_exclusivity_a}\\
e_i^{L+1} &\ge e_{\ell}^{L} - Mo_{i\ell jL},
\label{eq:edge_exclusivity_b}\ \\ & \forall i,\ell\in U,; i\neq\ell,; 
\forall L\in\mathcal{L}\setminus{L_{\max}},;
\forall j\in J_L. \notag
\end{align}

If   $ o_{i\ell jL}=1 $ , UAS $U_i$ is scheduled before UAS $U_\ell$ at slot $j$. Accordingly, UAS $U_\ell$ may depart its assigned slot on the upper level and enter the incoming descent edge only after UAS $U_i$ has exited the destination slot. Conversely, when ($o_{i\ell jL}=0 $), the second inequality becomes active and enforces the opposite ordering. Since every UAS traverses the incoming edge continuously without holding, these constraints ensure that the incoming descent corridors associated with a destination slot are used sequentially. Therefore, at any time instant, at most one UAS can occupy an incoming edge leading to the same ring slot.

\subsubsection{Pad capacity}
The landing pad is treated as level 0 with a single-server resource, so only one UAS can use it at a time.

 A minimum separation is maintained between two consecutive pad operations. Mathematically,
 \begin{align}
s_0 &= h_0 + \Delta^{(0)}
\label{eq:pad_capacity_a}
\end{align}

for the successive landing between UAS $i$ and $\ell$, the arrival times are constrained using the big-M method and the ordering variable. Mathematically, 
 \begin{align}
t_i^{0} &\ge t_{\ell}^{0} + s_{0} - M\left(1-o_{i\ell}^{(0)}\right),
\label{eq:pad_capacity_b} 
\end{align}
 \begin{align} \label{eq:pad_capacity_c}
t_{\ell}^{0} &\ge t_i^{0} + s_{0} - M\,o_{i\ell}^{(0)}, \\
\quad & \forall i,\ell \in U,\; i \neq \ell \notag
\end{align}

In the landing order between two UASs, one UAS lands before the other, mathematically,

\begin{align}
o_{i\ell}^{(0)} + o_{\ell i}^{(0)} &= 1, 
\quad \forall i,\ell \in U,\; i \neq \ell
\label{eq:pad_capacity_d}
\end{align}

\subsubsection{Spatial feasibility}
This constraint restricts the inner-ring slot choice to those that are geometrically feasible with the selected outer-ring slot. It preserves the funnel structure by allowing transitions only within the prescribed $\pm \theta_{max}^\circ$ angular band. Mathematically,

\begin{align}
\Delta(\theta_1,\theta_2) &= \min\left( | \theta_1- \theta_2 |, 360^\circ-|\theta_1-\theta_2|\right)
\label{eq:spatial_feasibility_a}\\
f_{j_{L},j'_{L-1}} &=
\begin{cases}
1, & \text{if } \Delta\!\left(\theta_{j_L},\phi_{j'_{L-1}}\right)\le \theta_{max}^\circ,\\
0, & \text{otherwise},
\end{cases}\\
& \forall j\in J_L  , \forall j'\in J_{L-1} \notag
\label{eq:spatial_feasibility_b}
\end{align}

A UAS can enter the slot in lower layer only if it is spatially feasible or that the UAS is directly added to lower layer exogenously. Mathematically, 
\begin{align}
x_{i,j'_{L-1},1} &\le \sum_{j_L \in J_{L}} f_{j_L,j'_{L-1}} \,x_{i,j_L,L} + \beta_{j'_{L-1}}, \\
\quad & \forall i \in U,\; \forall j'_{L-1} \in J_{L-1} \notag
\label{eq:spatial_feasibility_c}
\end{align}

\subsubsection{Wind-adjusted transit-time parameterization}
The transit time between two successive levels is computed from the three-dimensional path length and the wind-adjusted effective ground speed. This preserves the monotone descent assumption while allowing the travel time to vary with the selected descent direction and the wind component along that path.

The Cartesian position of slot $j\in J_L$ at level $L$ is defined as
\begin{equation}
\mathbf{r}_{j,L}
=
\begin{bmatrix}
r_L\cos\theta_{j,L}\\
r_L\sin\theta_{j,L}\\
z_L
\end{bmatrix}.
\label{eq:slot_position_vector}
\end{equation}

For a transition from slot $j_L\in J_L$ to slot $j'_{L-1}\in J_{L-1}$, the corresponding path length is
\begin{equation}
d^{(L)}_{j_L \rightarrow j'_{L-1}}
=
\left\|
\mathbf{r}_{j'_{L-1},L-1}
-
\mathbf{r}_{j_L,L}
\right\|_2 .
\label{eq:path_length_wind}
\end{equation}

The unit direction vector along the descent path is
\begin{equation}
\mathbf{u}^{(L)}_{j_Lj'_{L-1}}
=
\frac{
\mathbf{r}_{j'_{L-1},L-1}
-
\mathbf{r}_{j_L,L}
}{
d^{(L)}_{j_L \rightarrow j'_{L-1}}
}.
\label{eq:path_unit_vector_wind}
\end{equation}

The constant wind velocity vector is written as
\begin{equation}
\mathbf{V}_{w}
=
\begin{bmatrix}
V_w\cos\psi_w\\
V_w\sin\psi_w\\
V_{w,z}
\end{bmatrix},
\label{eq:wind_velocity_vector}
\end{equation}
where $V_w$ is the wind speed, $\psi_w$ is the wind direction, and $V_{w,z}$ is the vertical wind component.

The effective ground speed along the selected descent path is
\begin{equation}
v^{\mathrm{eff},(L)}_{j_Lj'_{L-1}}
=
\max
\left\{
v_{\min},
v_{\mathrm{air}}^{(L)}
+
\mathbf{V}_{w}^{T}
\mathbf{u}^{(L)}_{j_Lj'_{L-1}}
\right\}.
\label{eq:effective_ground_speed_wind}
\end{equation}

Here, $\mathbf{V}_{w}^{T}\mathbf{u}^{(L)}_{j_Lj'_{L-1}}$ represents the component of wind velocity along the selected descent direction. A positive value corresponds to a tailwind and increases the effective ground speed, whereas a negative value corresponds to a headwind and decreases it. The lower bound $v_{\min}>0$ is introduced to ensure that the effective speed remains strictly positive under strong headwind conditions, thereby avoiding zero or negative transit times in the scheduling model.\\

Therefore, the wind-adjusted transit time is
\begin{equation}
\delta^{(L)}_{j_L \rightarrow j'_{L-1}}
=
\frac{
d^{(L)}_{j_L \rightarrow j'_{L-1}}
}{
v^{\mathrm{eff},(L)}_{j_Lj'_{L-1}}
},
\quad
\forall L \in \mathcal{L}\setminus\{0\},\;
\forall j_L\in J_L,\;
\forall j'_{L-1}\in J_{L-1}.
\label{eq:transit_time_parameterization}
\end{equation}

Since the slot geometry and wind field are known before scheduling, the values of $\delta^{(L)}_{j_L \rightarrow j'_{L-1}}$ are precomputed and used as parameters in the optimization model. Therefore, the wind-adjusted transit-time model does not change the mixed-integer linear structure of the formulation.\\

\subsubsection{Temporal bounds}

Let $T_{\mathrm{op}}^{\max}>0$ denote the maximum allowable operational time for any UAS. Then,


\begin{subequations}\label{eq:horizon_bounds}
Limits the maximum admissible operational time of each UAS by bounding the difference between touchdown time and earliest arrival time.

\begin{equation}
t_i^{0}-\mathrm{ETA}_i \le T_{\mathrm{op}}^{\max},
\quad \forall i\in U
\label{eq:horizon_bounds_a}
\end{equation}

Defines the UAS-specific upper time bound by adding the operational-time cap to the corresponding arrival time.

\begin{equation}
T_i^{\max} = T_{\mathrm{op}}^{\max}+\mathrm{ETA}_i,
\quad \forall i\in U
\label{eq:horizon_bounds_b}
\end{equation}

Bounds the entry time at each level between the UAS arrival time and the maximum allowable scheduling time.

\begin{equation}
\mathrm{ETA}_i \le t_i^{L} \le T_i^{\max},
\quad \forall i\in U,\; \forall L \in \mathcal{L}
\label{eq:horizon_bounds_c}
\end{equation}

Bounds the exit time at each level within the same feasible scheduling horizon.

\begin{equation}
\mathrm{ETA}_i \le e_i^{L} \le T_i^{\max},
\quad \forall i\in U,\; \forall L \in \mathcal{L}
\label{eq:horizon_bounds_d}
\end{equation}

Restricts the holding time at each level to be nonnegative and no greater than the corresponding maximum allowable time.

\begin{equation}
0 \le h_{i,L} \le T_i^{\max},
\quad \forall i\in U,\; \forall L \in \mathcal{L}
\label{eq:horizon_bounds_e}
\end{equation}

Activates the entry-time variable only when a slot is selected at that level; otherwise, the bound forces it to zero.

\begin{equation}
t_i^{L} \le T_i^{\max}\sum_{j\in J_L} x_{i,j,L},
\quad \forall i\in U,\; \forall L \in \mathcal{L}
\label{eq:horizon_bounds_f}
\end{equation}

Activates the exit-time variable only when a slot is selected at that level; otherwise, the bound forces it to zero.

\begin{equation}
e_i^{L} \le T_i^{\max}\sum_{j\in J_L} x_{i,j,L},
\quad \forall i\in U,\; \forall L \in \mathcal{L}
\label{eq:horizon_bounds_g}
\end{equation}

\end{subequations}

\subsection{Greedy Solution Approach}
\label{subsec:greedy_approach}

Solving the complete arrival-planning problem simultaneously for all UAS becomes computationally demanding as the number of vehicles, ring slots, and airspace levels increases. This limits the use of the centralized formulation in real-time operations, particularly when new arrivals, emergency landing requests, or wind disturbances require rapid replanning. To reduce the computational burden, the proposed greedy approach first establishes a planning order and subsequently solves the optimization problem for one UAS at a time. After a UAS is planned, its trajectory is fixed and treated as a space--time reservation when planning the remaining UAS.

A sequencing priority $\rho_i$ is introduced specifically for the greedy solution procedure. This parameter is distinct from the emergency weight $\alpha_i$ used in the objective function. The parameter $\rho_i$ determines when UAS $i$ is planned, whereas $\alpha_i$ influences the trajectory and landing time selected by its optimization subproblem. A larger value of $\rho_i$ indicates that the UAS should be planned earlier.

Let
\begin{equation}
\boldsymbol{\pi}
=
(\pi_1,\pi_2,\ldots,\pi_N)
\end{equation}
denote the ordered UAS sequence, where $\pi_k\in\mathcal{U}$ is the UAS considered at iteration $k$. The sequence is constructed using $\rho_i$ as the primary criterion and $\mathrm{ETA}_i$ as the secondary criterion. Thus, the precedence relation is defined as
\begin{equation}
i\prec \ell
\quad\Longleftrightarrow\quad
\rho_i>\rho_\ell
\quad\lor\quad
\left(
\rho_i=\rho_\ell
\land
\mathrm{ETA}_i<\mathrm{ETA}_\ell
\right),
\label{eq:greedy_order}
\end{equation}
where $i\prec\ell$ indicates that UAS $i$ is planned before UAS $\ell$. Emergency UAS are assigned higher sequencing priorities, while UAS within the same priority class follow a first-come-first-served order based on $\mathrm{ETA}_i$. A fixed UAS identifier may be used to resolve any remaining ties.

At iteration $k$, the set of previously planned UAS is
\begin{equation}
\mathcal{U}^{\mathrm{fix}}_k
=
\{\pi_1,\pi_2,\ldots,\pi_{k-1}\}.
\label{eq:greedy_fixed_set}
\end{equation}
Only the decision variables associated with UAS $\pi_k$ are optimized:
\begin{equation}
\left\{
x_{\pi_k,j,L},
y^{(L)}_{\pi_k,j_L,j'_{L-1}},
t_{\pi_k}^{L},
e_{\pi_k}^{L},
h_{\pi_k,L}
\right\}.
\label{eq:greedy_decision_variables}
\end{equation}
The slot assignments, selected edges, and timing variables of every UAS $\ell\in\mathcal{U}^{\mathrm{fix}}_k$ remain fixed. The optimization model for UAS $\pi_k$ retains the original objective function, including its emergency weight $\alpha_{\pi_k}$, and satisfies the assignment, angular-feasibility, connectivity, travel-time, holding-time, slot-capacity, edge-exclusivity, and landing-pad separation constraints.

The fixed trajectories define occupied time intervals for the corresponding ring slots, connecting edges, and landing pad. A candidate resource can therefore be assigned to UAS $\pi_k$ only if its occupancy interval satisfies the required headway and does not overlap any previously reserved interval. If UAS $\pi_k$ uses slot $j\in J_L$ after a previously planned UAS $\ell$, then
\begin{equation}
t_{\pi_k}^{L}
\geq
e_{\ell}^{L}+\Delta^{(L)},
\label{eq:greedy_slot_separation}
\end{equation}
and the associated ordering variable is fixed consistently as $o_{\pi_k\ell jL}=1$. Equivalent reservation conditions are imposed when the two UAS share a connecting edge or the landing pad. Once a feasible plan is obtained for UAS $\pi_k$, its decision variables are fixed before UAS $\pi_{k+1}$ is considered.

\begin{algorithm}[tbp]
\caption{Priority-aware greedy UAS arrival planning}
\label{alg:greedy_planning}
\begin{algorithmic}[1]
\Require UAS set $\mathcal{U}$, sequencing priorities $\rho_i$, emergency weights $\alpha_i$, entry times $\mathrm{ETA}_i$, slot sets $J_L$, angular-feasibility parameters $f_{j_L,j'_{L-1}}$, travel times $\delta^{(L)}_{j_L\to j'_{L-1}}$, and safety headways $\Delta^{(L)}$
\Ensure Conflict-free slot assignments, paths, and timing decisions

\State Construct $\boldsymbol{\pi}$ by sorting $\mathcal{U}$ by decreasing $\rho_i$ and increasing $\mathrm{ETA}_i$
\State $\mathcal{U}^{\mathrm{fix}}\gets\emptyset$

\For{$k=1$ to $N$}
    \State $i\gets\pi_k$

    \State Solve the optimization problem for UAS $i$  subject to  fixed trajectories $\mathcal{U}^{\mathrm{fix}}$  
    
    \State Fix the slot assignments, selected edges, and timing decisions of UAS $i$
    \State $\mathcal{U}^{\mathrm{fix}} \gets \mathcal{U}^{\mathrm{fix}}\cup\{i\}$
    
\EndFor

\State \Return the planned trajectories of all UAS in $\mathcal{U}^{\mathrm{fix}}$
\end{algorithmic}
\end{algorithm}

The complete greedy procedure is summarized in Algorithm~\ref{alg:greedy_planning}. Within the rolling-horizon implementation, Algorithm~\ref{alg:greedy_planning} is repeated whenever new traffic information or an operational disturbance becomes available. Newly arriving UAS are inserted into the sequence according to $\rho_i$ and $\mathrm{ETA}_i$, while completed trajectory segments remain fixed. When an emergency landing request is received, the affected UAS is assigned a higher sequencing priority $\rho_i$ and moved forward in the replannable sequence. Its emergency weight $\alpha_i$ remains in the objective function to favor an expedited landing trajectory. Following a wind disturbance, the affected travel times $\delta^{(L)}_{j_L\to j'_{L-1}}$ are updated using the wind-adjusted effective speeds $v^{\mathrm{eff},(L)}_{j_Lj'_{L-1}}$, and only the affected future trajectories are replanned.

The proposed method converts the original coupled optimization problem into a sequence of smaller single-UAS problems. Fixing the planning order eliminates the need to jointly optimize the pairwise sequencing decisions of all UAS, thereby reducing the problem size and computation time. Although the greedy procedure does not guarantee a globally optimal solution because previously fixed trajectories are not reconsidered, it preserves conflict-free resource usage and provides the computational efficiency required for real-time UAS arrival planning.

\section{Results}
\subsection{Simulation Setup}
The terminal airspace follows Fig.~1 with two holding rings above the
landing pad. The outer ring has a radius of 500\,m at 200\,m altitude and
eight slots (N, NE, E, SE, S, SW, W, NW). The inner ring has a radius of
200\,m at 100\,m altitude and four slots (N, E, S, W). UAS enter at a radius
of 800\,m and an altitude of 500\,m and descend ring by ring to the pad at the
centre. The minimum holding time is 10\,s on each ring, the slot headway is
15\,s, and the pad clearance between consecutive landings is 30\,s. Without
wind, the transit times are 52.7\,s (outer to inner ring) and 55.9\,s (inner
ring to pad). The case studies use a fixed list of 12~UAS with the same
arrival times and sectors in all four cases. In the Monte-Carlo studies (no
wind, equal priority), arrivals follow a Poisson process with rate
$\lambda = 1/T$ over 3600\,s, averaged over 300 replications. In the
arrival-rate study, $T \in \{60, 75, 90, 120, 150, 180\}$\,s with uniformly
distributed entry sectors. In the arrival-direction study, $\lambda$ is fixed
(one UAS per 78.4\,s) and the arrival bearing follows a von Mises distribution
centred on north, with $\kappa \in \{0, 0.5, 1, 2, 4, 8\}$.

\subsection{Computational Setup and Overview of Results}
All experiments were run in Python~3.10 with Gurobi~13.0.2 (university
license) on a computer with an Intel Core~i7 processor and 16\,GB of RAM.
Section~IV-C shows the qualitative flow of the UAS through the rings using
snapshots of Case~IV. Section~IV-D compares four case studies on the 12-UAS
problem: Case~I (equal priority, no wind), Case~II (priority, no wind),
Case~III (equal priority, wind) and Case~IV (priority and wind).
Section~IV-E presents the ablation studies on the arrival rate, the number of
rings, and the spatial distribution of arrivals.

\subsection{Poisson Entry of UASs with Wind Disturbance and Priority}

\begin{figure*}[t]
    \centering
    \setlength{\tabcolsep}{2pt}
    \renewcommand{\arraystretch}{1.0}
    \begin{tabular}{ccc}
        \subfloat[$t = 430\,\mathrm{s}$\label{fig:snap1}]{
            \includegraphics[width=0.30\textwidth]{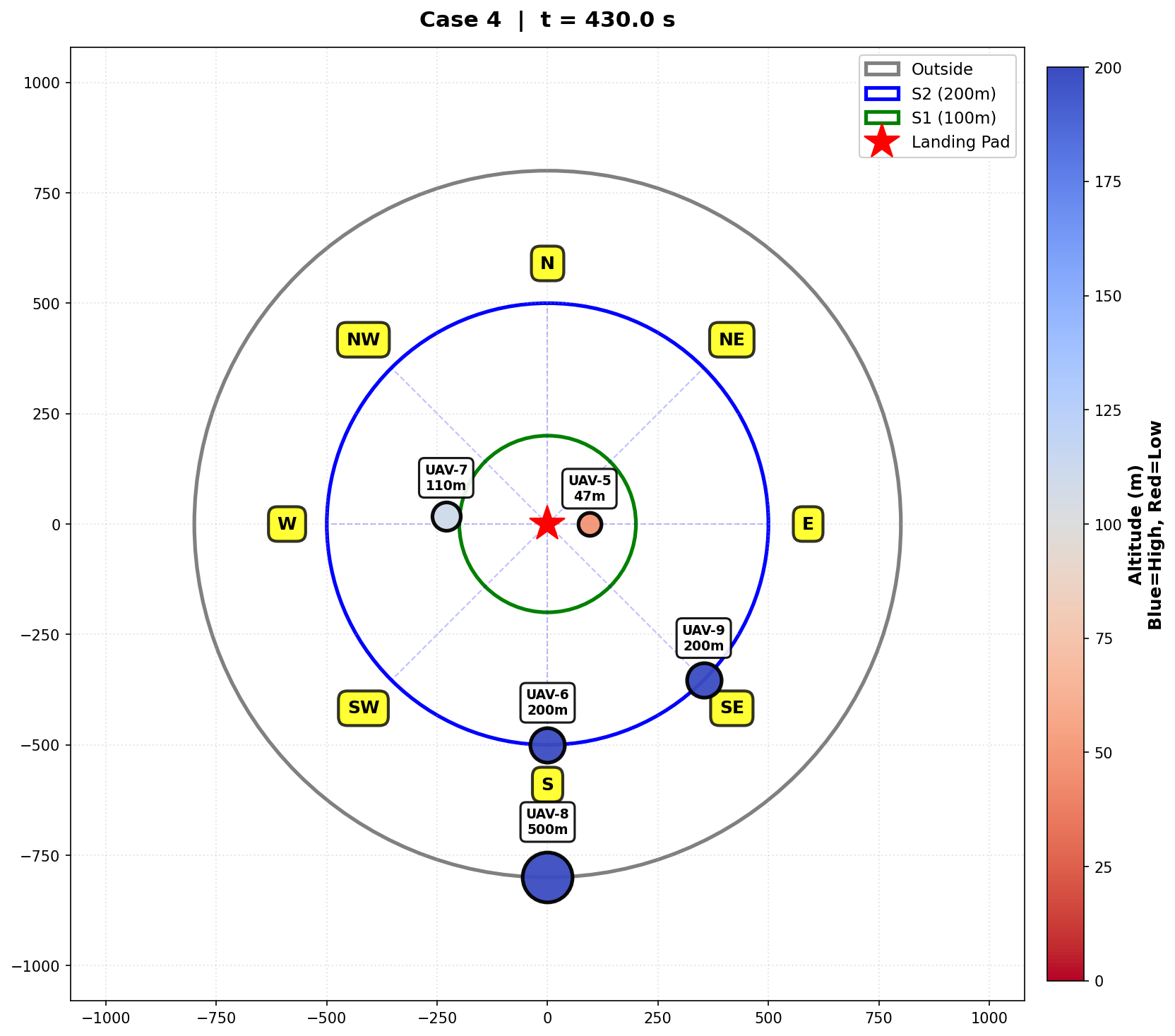}
        } &
        \subfloat[$t = 435\,\mathrm{s}$\label{fig:snap2}]{
            \includegraphics[width=0.30\textwidth]{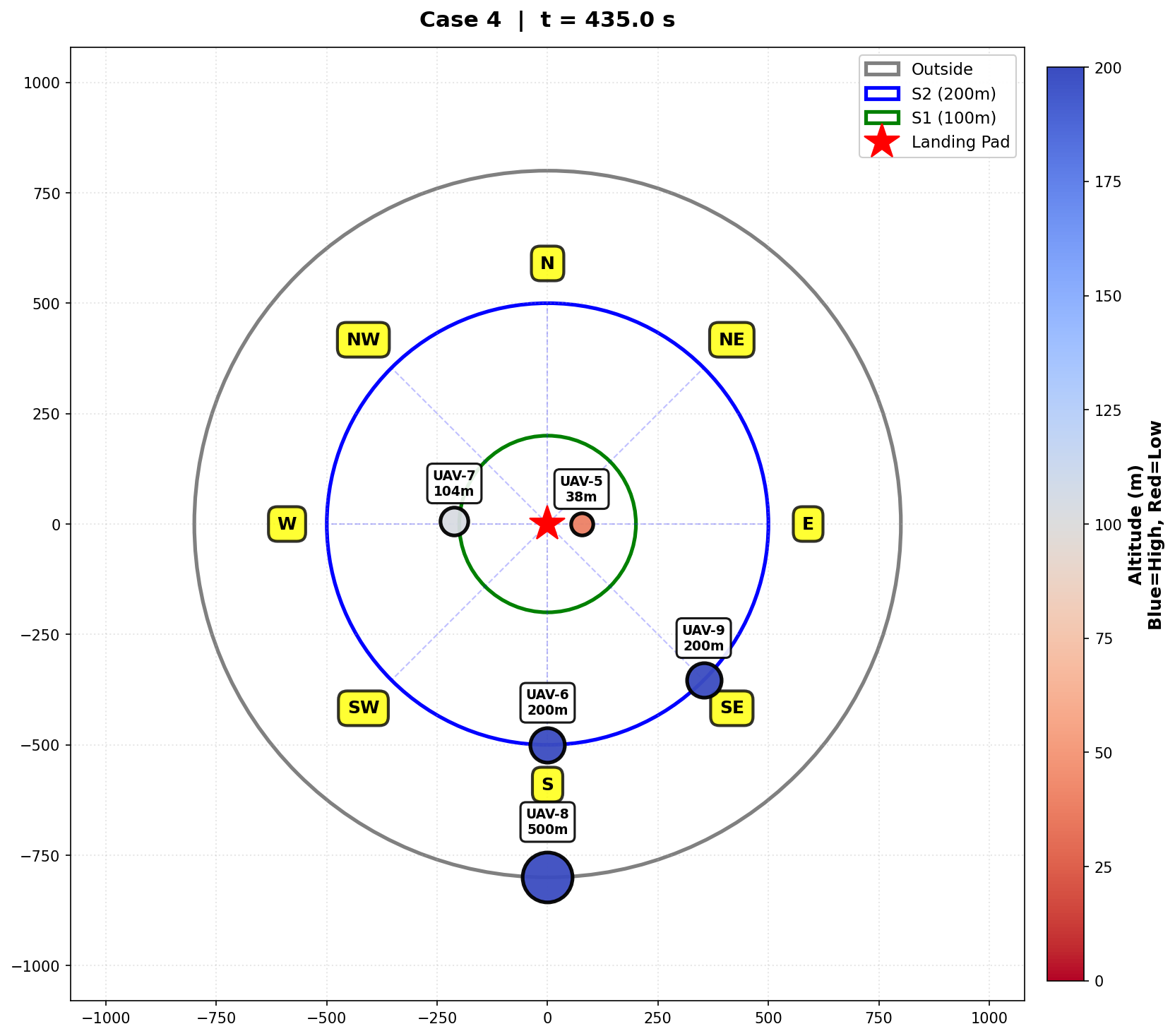}
        } &
        \subfloat[$t = 471.2\,\mathrm{s}$\label{fig:snap3}]{
            \includegraphics[width=0.30\textwidth]{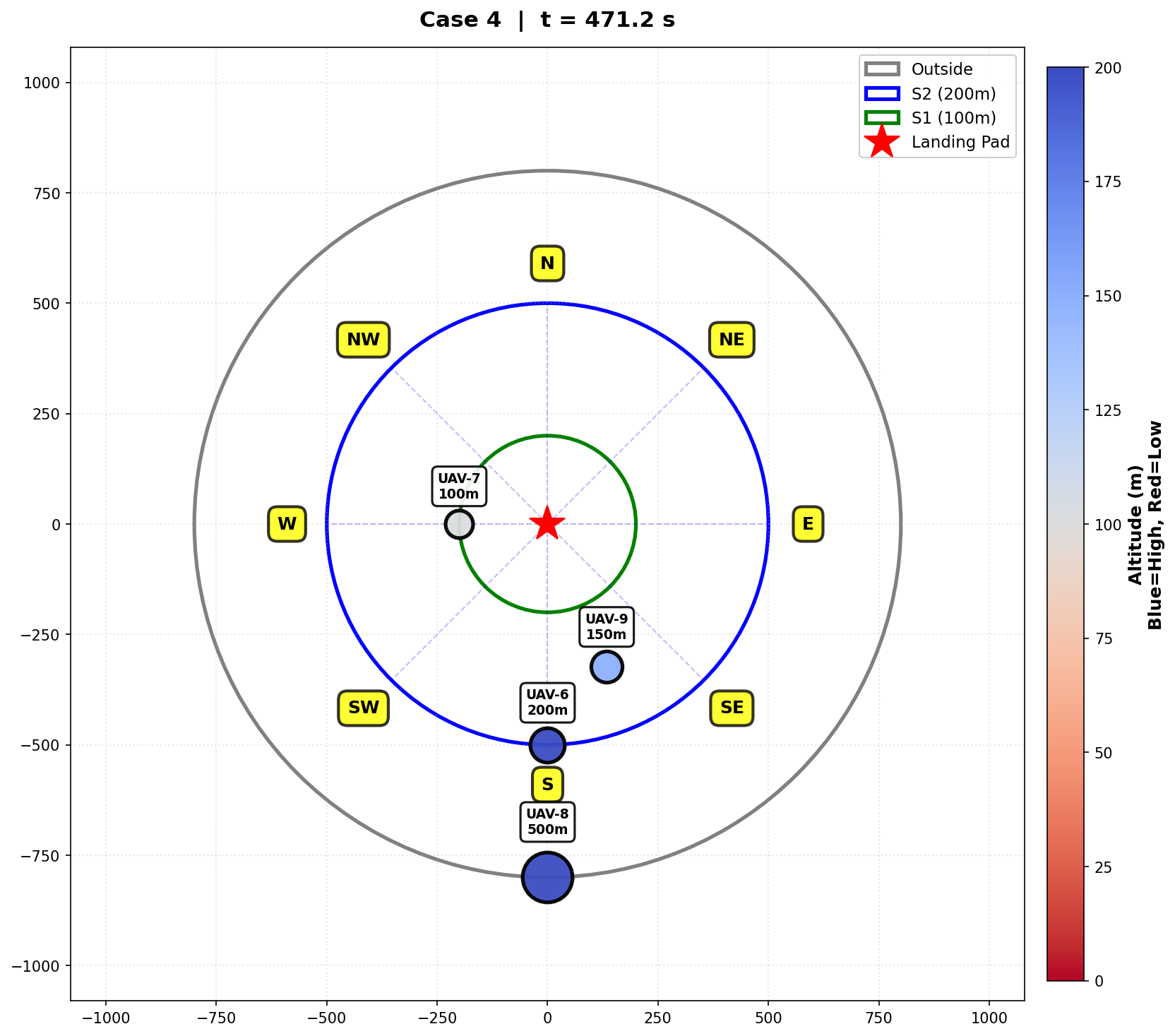}
        } \\

        \subfloat[$t = 535.2\,\mathrm{s}$\label{fig:snap4}]{
            \includegraphics[width=0.30\textwidth]{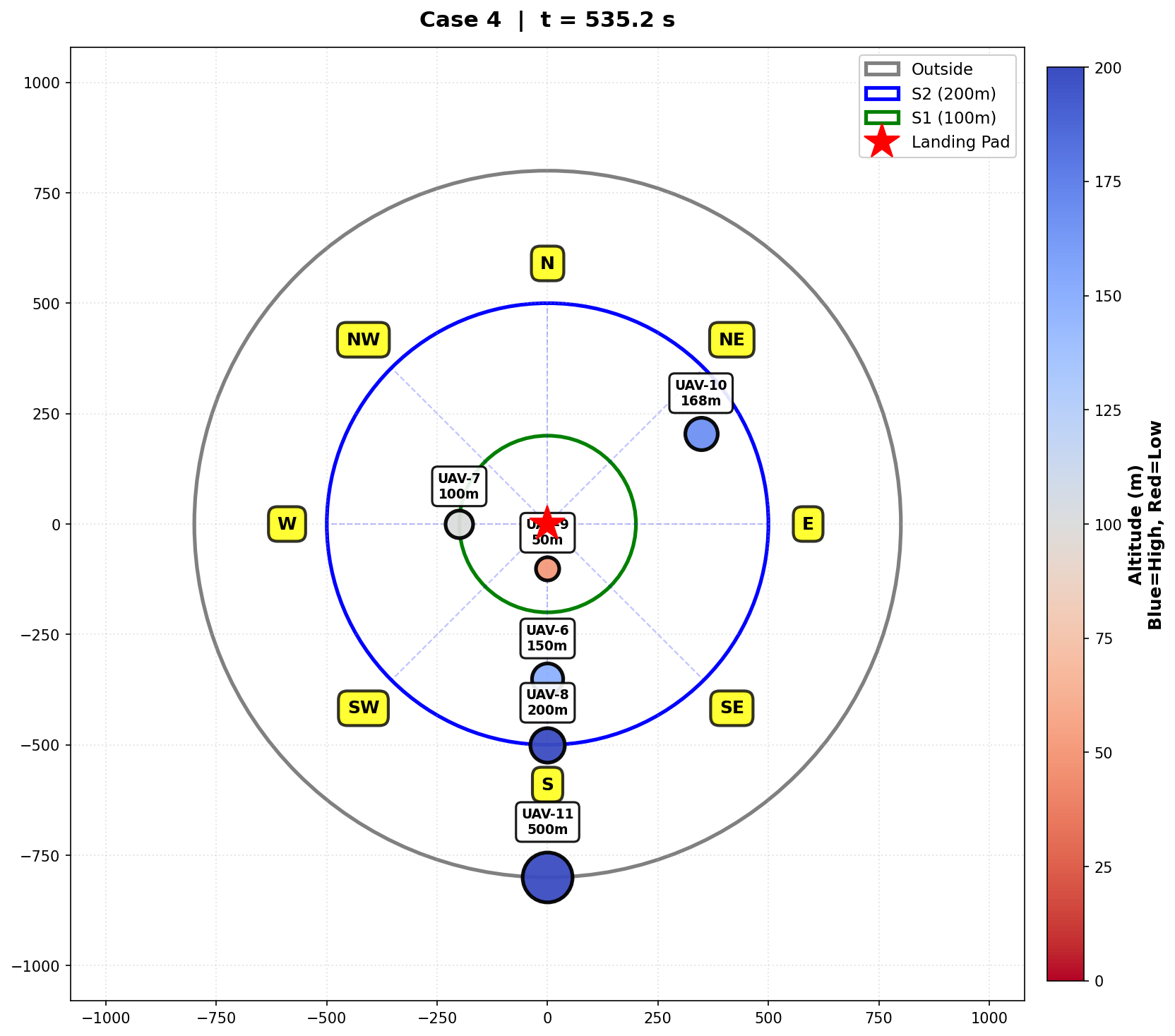}
        } &
        \subfloat[$t = 558\,\mathrm{s}$\label{fig:snap5}]{
            \includegraphics[width=0.30\textwidth]{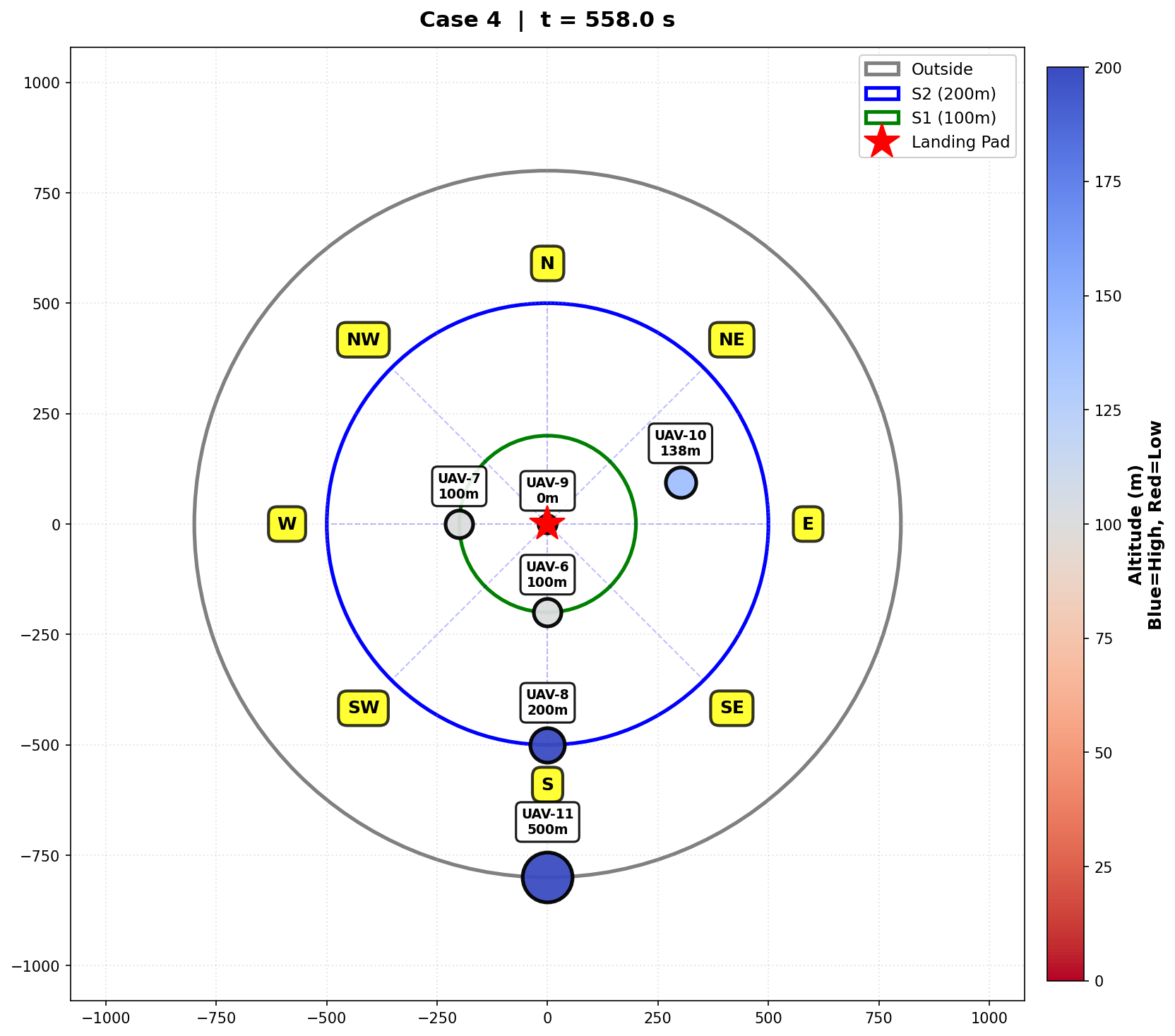}
        } &
        \subfloat[$t = 613.9\,\mathrm{s}$\label{fig:snap6}]{
            \includegraphics[width=0.30\textwidth]{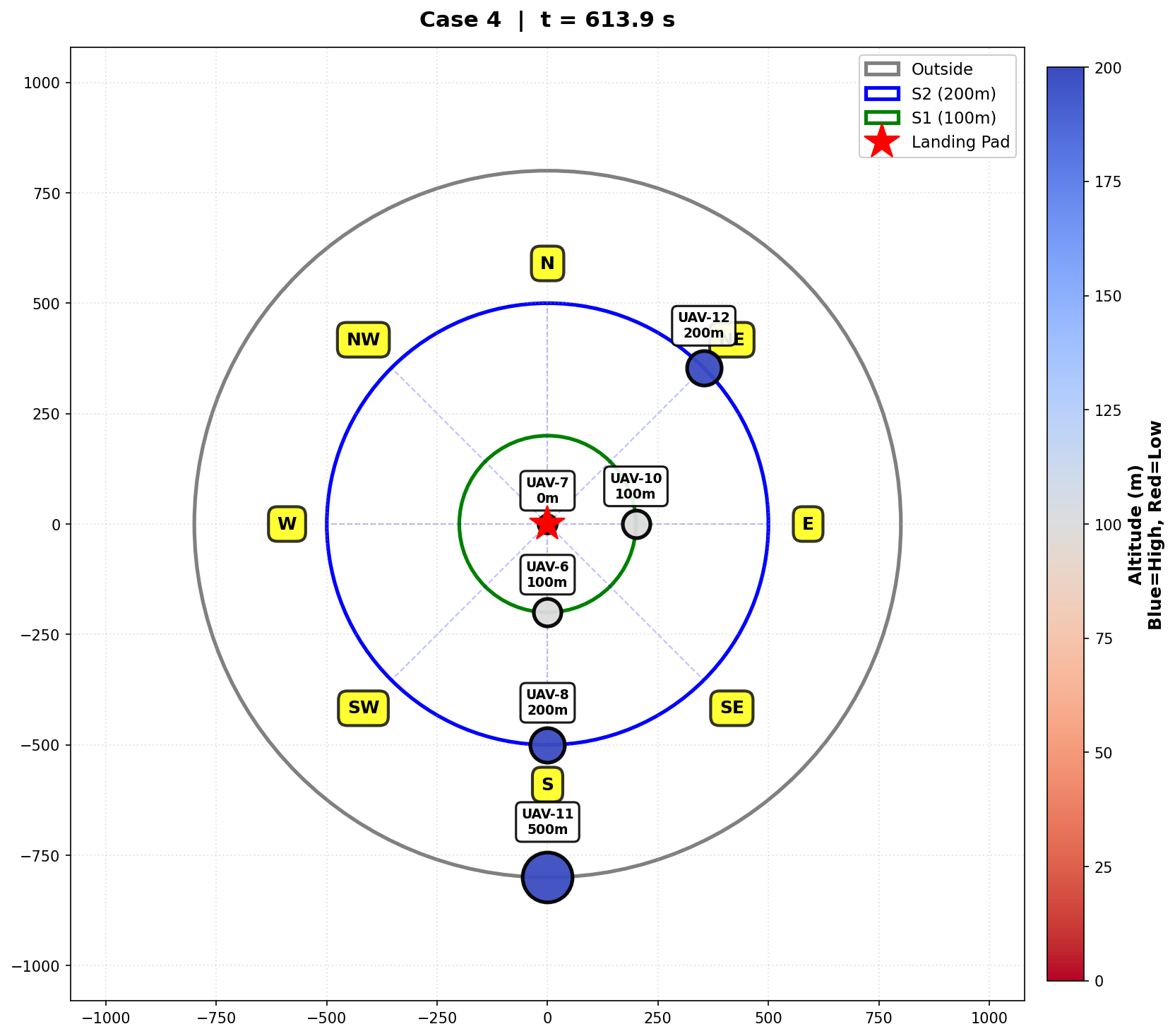}
        }
    \end{tabular}
    \caption{Representative top-view snapshots of Case~IV, in which both the priority
request and the wind disturbance are considered, showing the UAS movements at
different time instants. }
    \label{fig:result_snapshots}
\end{figure*}

Fig.~\ref{fig:result_snapshots} shows representative snapshots of the landing sequence for Case~4, where UAS-9 has a priority weight of $\alpha = 5$ and a wind of $1~\mathrm{m/s}$ blows toward the north. At $t = 430~\mathrm{s}$, UAS-9 enters outer slot SE and holds for only the minimum $10~\mathrm{s}$, while UAS-6 and UAS-7, which arrived earlier, are still holding. By $t = 535.2~\mathrm{s}$, UAS-9 is already on its final descent to the pad, and at $t = 558~\mathrm{s}$ it touches down with a delay of $128.05~\mathrm{s}$, ahead of UAS-6, UAS-7 and UAS-8. UAS-7 leaves its inner slot at the same instant and lands at $t = 613.9~\mathrm{s}$, $55.9~\mathrm{s}$ later. The descents from the southern sectors are also shorter under the northward wind than the NE to E transfer. The snapshots are consistent with the slot-capacity and pad-separation constraints.

\subsection{Case Studies} 
\begin{figure*}[htbp]
\centering
\subfloat[Case 1]{%
    \includegraphics[width=0.48\linewidth]{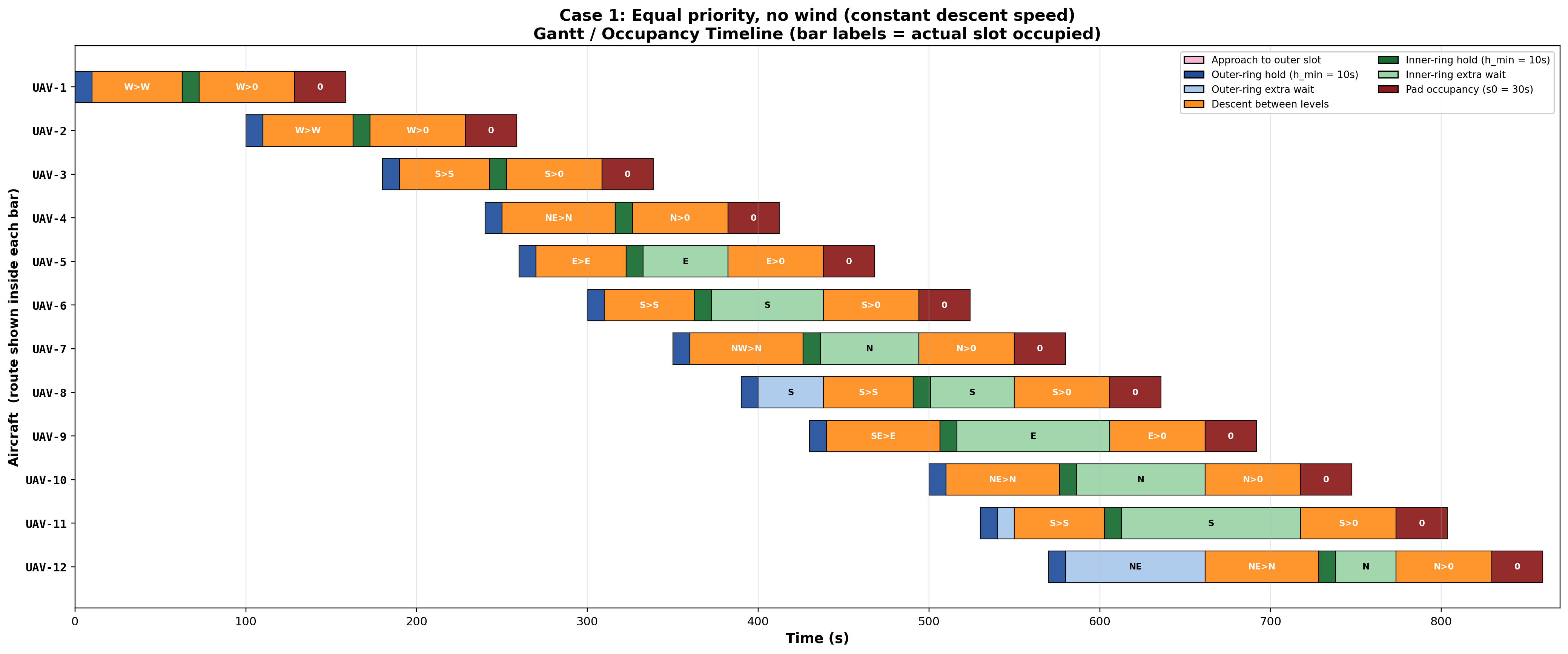}
    \label{fig:case1}
}
\subfloat[Case 2]{%
    \includegraphics[width=0.48\linewidth]{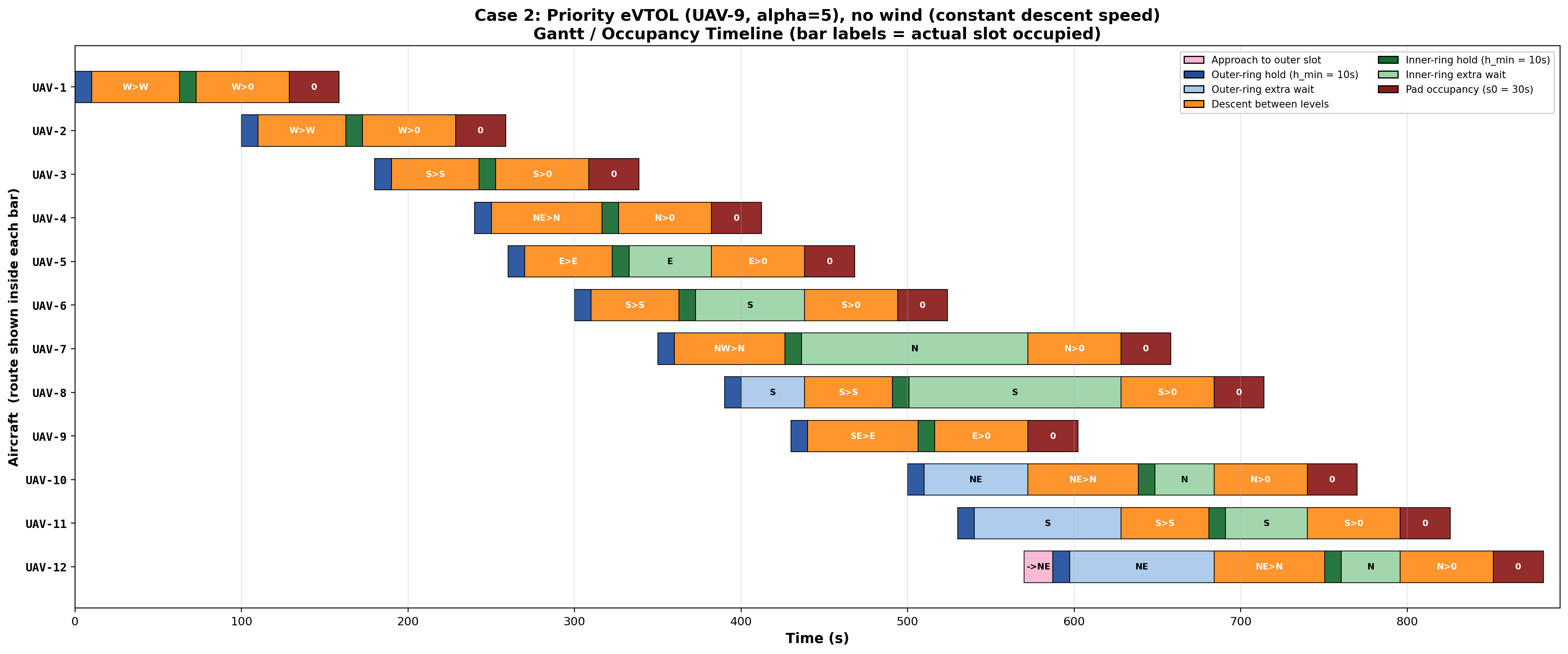}
    \label{fig:case2}
} \\
\subfloat[Case 3]{%
    \includegraphics[width=0.48\linewidth]{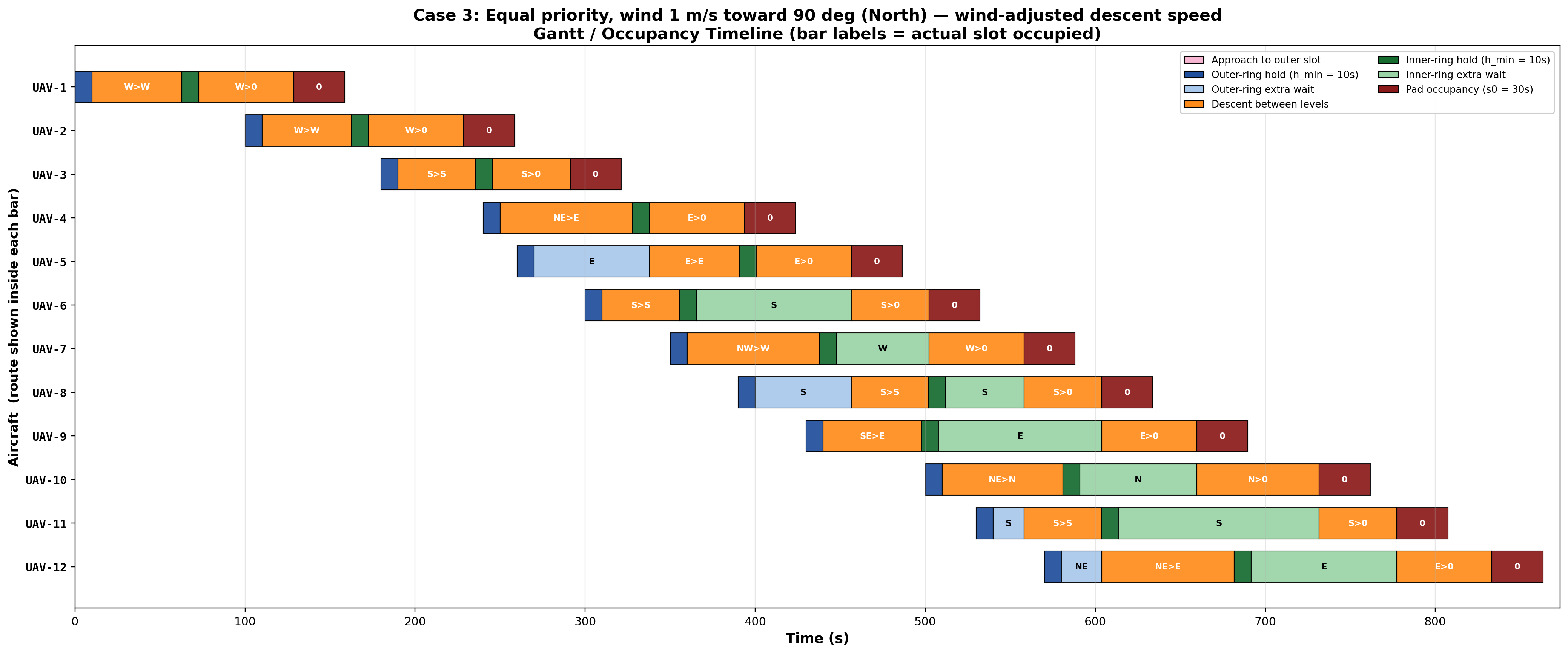}
    \label{fig:case3}
}
\subfloat[Case 4]{%
    \includegraphics[width=0.48\linewidth]{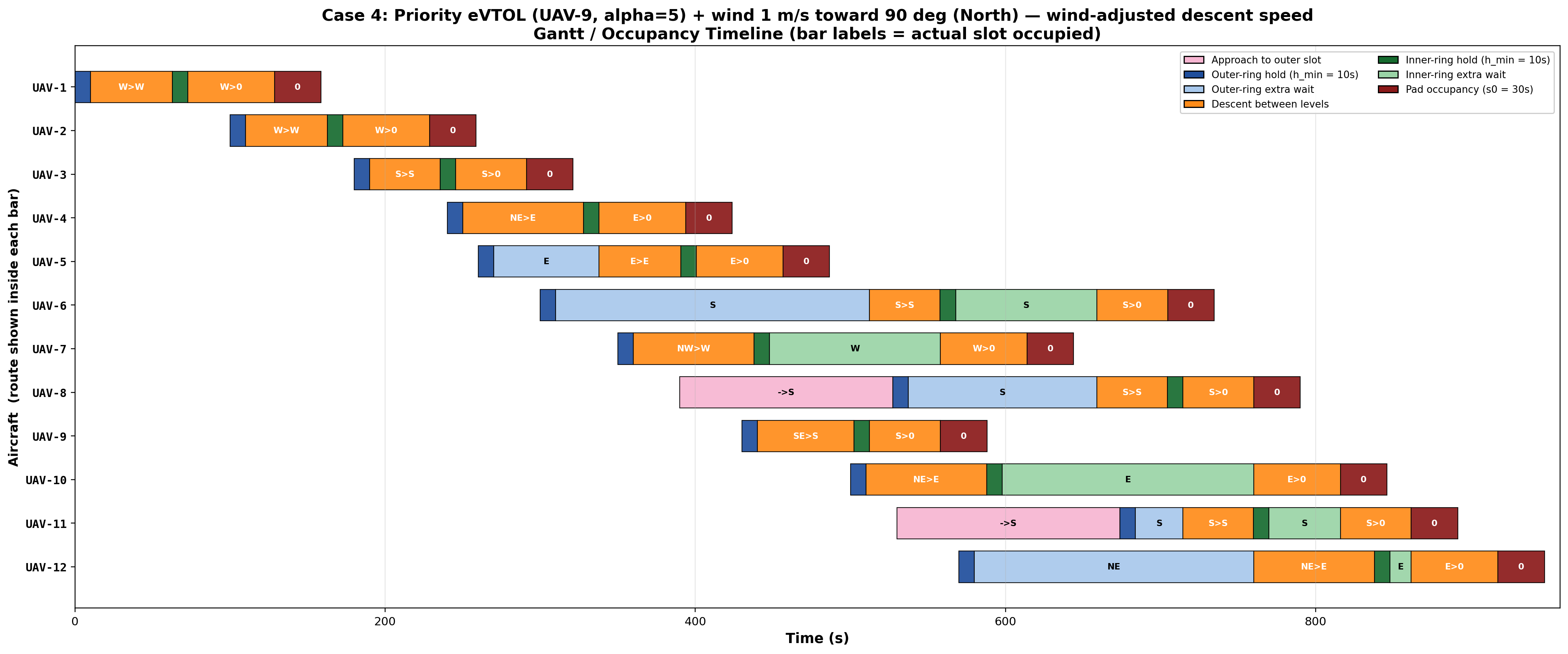}
    \label{fig:case4}
}
\caption{Gantt charts for UAS landing.}
\label{fig:gantt}
\end{figure*}

The features of the proposed method are demonstrated using 4 cases in a 12-UAS landing problem. All cases have the same arrival times and arrival locations of UASs. The landing schedules (gantt chart) are shown in fig. \ref{fig:gantt}.   Table \ref{table:tb1} shows the landing sequence, the path of UAS in rings (outer ring to inner ring to landing), landing time and the operational time of UASs (time between arrival and landing).

A detailed video of these case studies is available at \href{https://youtu.be/F9IP1dz-9PE}{https://youtu.be/F9IP1dz-9PE}

\begin{table*}[htb]
\centering
\caption{Comparison of the four cases} \label{table:tb1}
\resizebox{\textwidth}{!}{%
\begin{tabular}{|l|l|r|r|l|l|r|r|l|l|r|r|l|l|r|r|}
\hline
\multicolumn{4}{|c|}{\textbf{Case 1}} &
\multicolumn{4}{c|}{\textbf{Case 2}} &
\multicolumn{4}{c|}{\textbf{Case 3}} &
\multicolumn{4}{c|}{\textbf{Case 4}} \\
\hline
{\begin{tabular}[c]{@{}c@{}} UAS \\ Id \end{tabular}}   & Path 
& {\begin{tabular}[c]{@{}c@{}}Landing \\ Time\\ (sec)\end{tabular}}  
& {\begin{tabular}[c]{@{}c@{}} Operational \\ Time\\ (sec)\end{tabular}}   
& {\begin{tabular}[c]{@{}c@{}} UAS \\ Id \end{tabular}} & Path 
& {\begin{tabular}[c]{@{}c@{}}Landing \\ Time\\ (sec)\end{tabular}}  
& {\begin{tabular}[c]{@{}c@{}} Operational \\ Time\\ (sec)\end{tabular}}  
& {\begin{tabular}[c]{@{}c@{}} UAS \\ Id \end{tabular}} & Path 
& {\begin{tabular}[c]{@{}c@{}}Landing \\ Time\\ (sec)\end{tabular}}  
&  {\begin{tabular}[c]{@{}c@{}} Operational \\ Time\\ (sec)\end{tabular}}   
& {\begin{tabular}[c]{@{}c@{}} UAS \\ Id \end{tabular}} & Path 
& {\begin{tabular}[c]{@{}c@{}}Landing \\ Time\\ (sec)\end{tabular}}  
& {\begin{tabular}[c]{@{}c@{}} Operational \\ Time\\ (sec)\end{tabular}}   \\
\hline
UAS1  & W-W-0  & 128.61 & 128.61 & 1  & W-W-0  & 128.61 & 128.61 & 1  & W-W-0  & 128.61 & 128.61 & 1  & W-W-0  & 128.61 & 128.61 \\
UAS2  & W-W-0  & 128.61 & 128.61 & 2  & W-W-0  & 228.61 & 128.61 & 2  & W-W-0  & 228.61 & 128.61 & 2  & W-W-0  & 228.61 & 128.61 \\
UAS3  & S-S-0  & 128.61 & 128.61 & 3  & S-S-0  & 308.61 & 128.61 & 3  & S-S-0  & 291.20 & 111.20 & 3  & S-S-0  & 291.20 & 111.20 \\
UAS4  & NE-N-0 & 142.27 & 142.27 & 4  & NE-N-0 & 382.27 & 142.27 & 4  & NE-E-0 & 393.80 & 153.80 & 4  & NE-E-0 & 393.80 & 153.80 \\
UAS5  & E-E-0  & 178.17 & 178.17 & 5  & E-E-0  & 438.17 & 178.17 & 5  & E-E-0  & 456.50 & 196.50 & 5  & E-E-0  & 456.50 & 196.50 \\
UAS6  & S-S-0  & 194.07 & 194.07 & 6  & S-S-0  & 494.07 & 194.07 & 6  & S-S-0  & 502.19 & 202.19 & 9  & SE-S-0 & 558.05 & 128.05 \\
UAS7  & NW-N-0 & 199.98 & 199.98 & 9  & SE-E-0 & 572.27 & 142.27 & 7  & NW-W-0 & 558.09 & 208.09 & 7  & NW-W-0 & 613.95 & 263.95 \\
UAS8  & S-S-0  & 605.88 & 215.88 & 7  & NW-N-0 & 628.17 & 278.17 & 8  & S-S-0  & 603.78 & 213.78 & 6  & S-S-0  & 704.55 & 404.55 \\
UAS9  & SE-E-0 & 661.78 & 231.78 & 8  & S-S-0  & 684.07 & 294.07 & 9  & SE-E-0 & 659.68 & 229.68 & 8  & S-S-0  & 760.06 & 370.06 \\
UAS10 & NE-N-0 & 717.68 & 217.68 & 10 & NE-N-0 & 739.98 & 239.98 & 10 & NE-N-0 & 731.68 & 231.68 & 10 & NE-E-0 & 815.96 & 315.96 \\
UAS11 & S-S-0  & 773.58 & 243.58 & 11 & S-S-0  & 795.88 & 265.88 & 11 & S-S-0  & 777.37 & 247.37 & 11 & S-S-0  & 861.65 & 331.65 \\
UAS12 & NE-N-0 & 829.49 & 259.49 & 12 & NE-N-0 & 851.78 & 281.78 & 12 & NE-E-0 & 833.27 & 263.27 & 12 & NE-E-0 & 917.55 & 347.55 \\
\hline
\end{tabular}
}
\end{table*}

\subsubsection{Case I } Equal priority of all UASs and no wind disturbance

The first case shows the landing sequence for all 12 UASs in \ref{fig:case1}. UASs land in the same sequence as they arrive. The $U5$ enters the ring 1 from east direction, then it wait   until  $U4$ lands from north direction. Only after that the $U5$ moves for landing.   The average operational time is 189.0604 secs.

\subsubsection{Case II}:  Priority and no wind disturbance. 

To illustrate the time-critical landing of UASs, UAS $U9$ is set to higher priority ($\alpha_9 = 5$), all remaining UASs has unit priority. The gantt chart is shown in fig. \ref{fig:case2}. Due to the priority, $U9$ needs to land with minimum delay. One can observe that,$U9$ lands in the minimum operational time without any delay. UASs $U7$ and $U8$ are delayed to prioritize landing of $U9$. As a consequence, the overall delay increases in systems and the normalised operational time increases to 247.6317 secs.

\subsubsection{Case III} Equal priority and wind disturbance.
The ability to handle wind disturbance are demonstrated with wind speed of $1 m/s$ in north direction. All UASs land in a sequence same as the case I. But they change their routes. The $U4$ entering from NE, now chooses east direction for ring 1 for landing, to avoid the opposite wind. Due to this $U5$ entering from east needs to wait and the delay propagates in system. From the table \ref{table:tb1}, one can observe that the $U3$ coming from south, traveling with wind direction has lower operational time. The UASs traveling in opposite directions require extra time to overcome the winds. The overall landing time changes and is reported in table. The average operational tine is 192.8963 secs.
 
 \subsubsection{Case IV}  Priority and wind disturbance. 

 UAS 9 is set to higher priority in a  wind flowing in the North direction  with speed $1 m/s$. 
 Combined effect increases the normalized operational time to 282.7223 sec. $U9$ lands without any dealy with minimum operational time after $U5$. $U6,U7, U8$ are kept on waiting for sometime to provide the freeway for $U9$. Due to wind, landing sequence for the remaining UASs get updated and it does not follow FIFO policy.

\subsection{Ablation studies}
The performance of the proposed method is evaluated using Monte-Carlo simulations for different parameters. The arrivals of UASs is quite unknown in real time, and we study the effect of arrival rate by varying the Poisson distribution rate parameter and the spatial distribution by changing the arrival positions using the Von Mises distribution.

\begin{figure}[t]
    \centering
    \includegraphics[width=\linewidth]{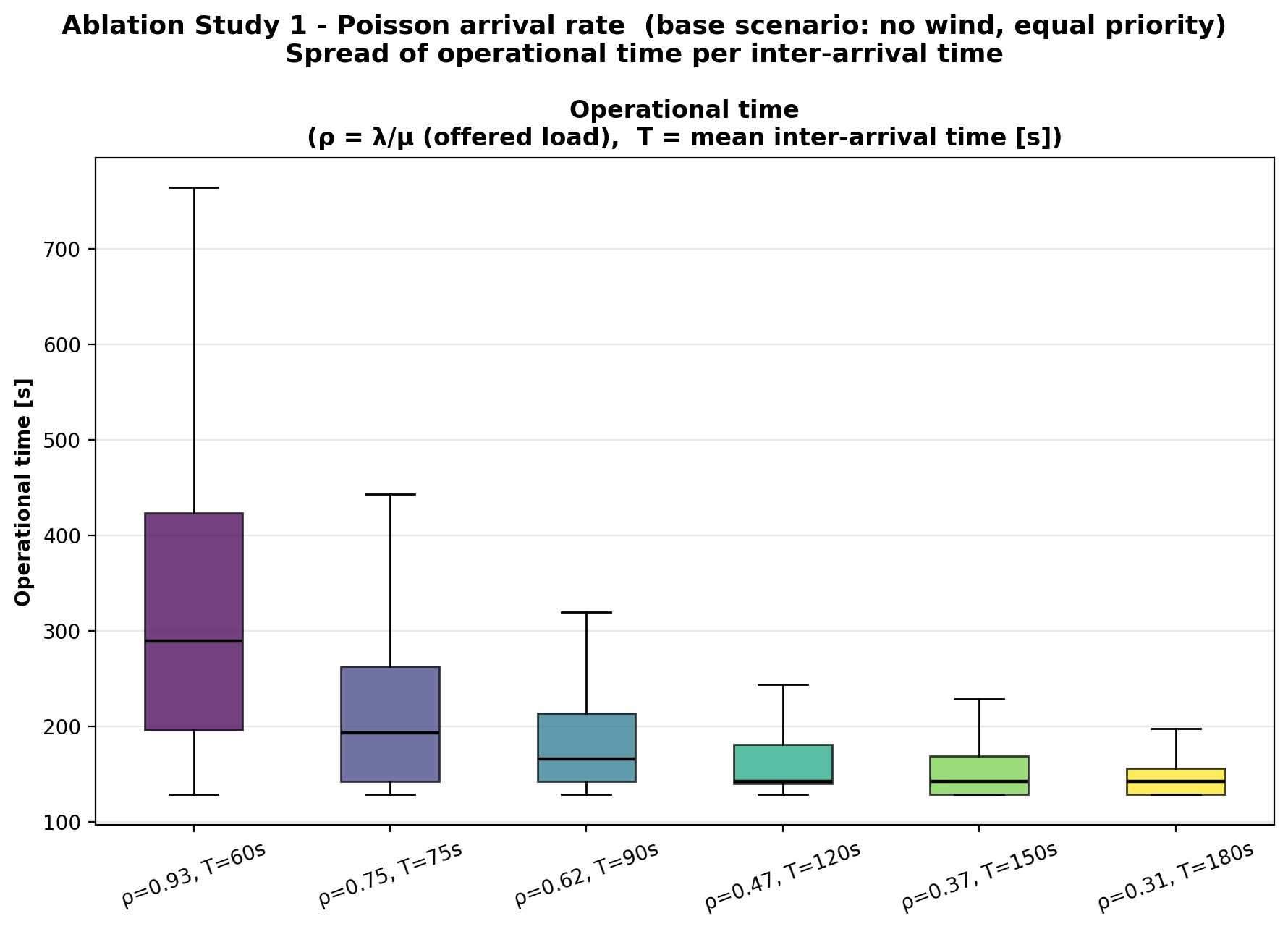}
    \caption{Effects of different arrival load }
    \label{fig:operational_time_poison}
\end{figure}

\subsubsection{Effect of Arrival Rate}

\begin{figure*}[htbp]
\centering
\subfloat[]{%
    \includegraphics[width=0.32\linewidth]{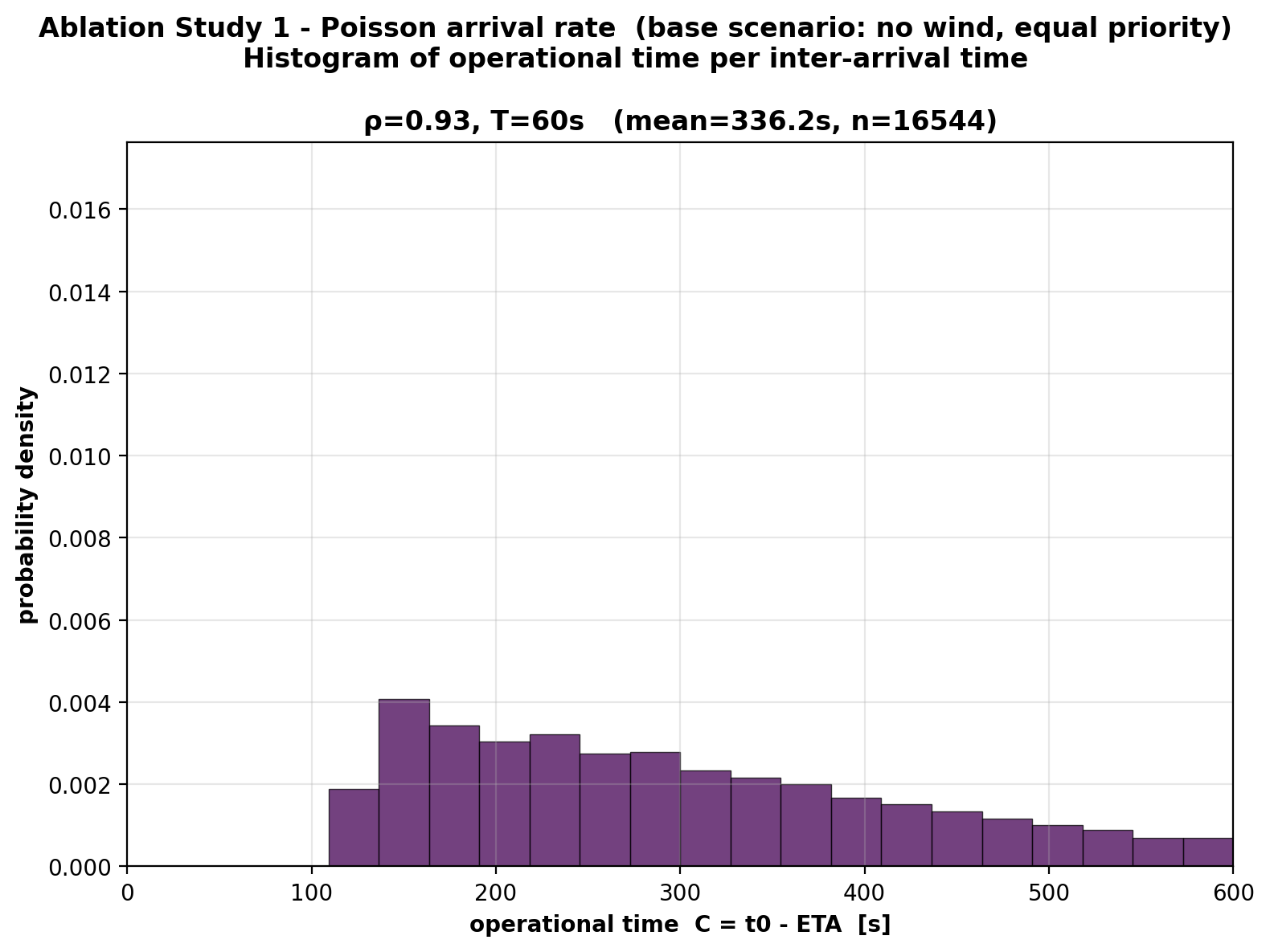}
    \label{fig:histohram1}
}
\subfloat[]{%
    \includegraphics[width=0.32\linewidth]{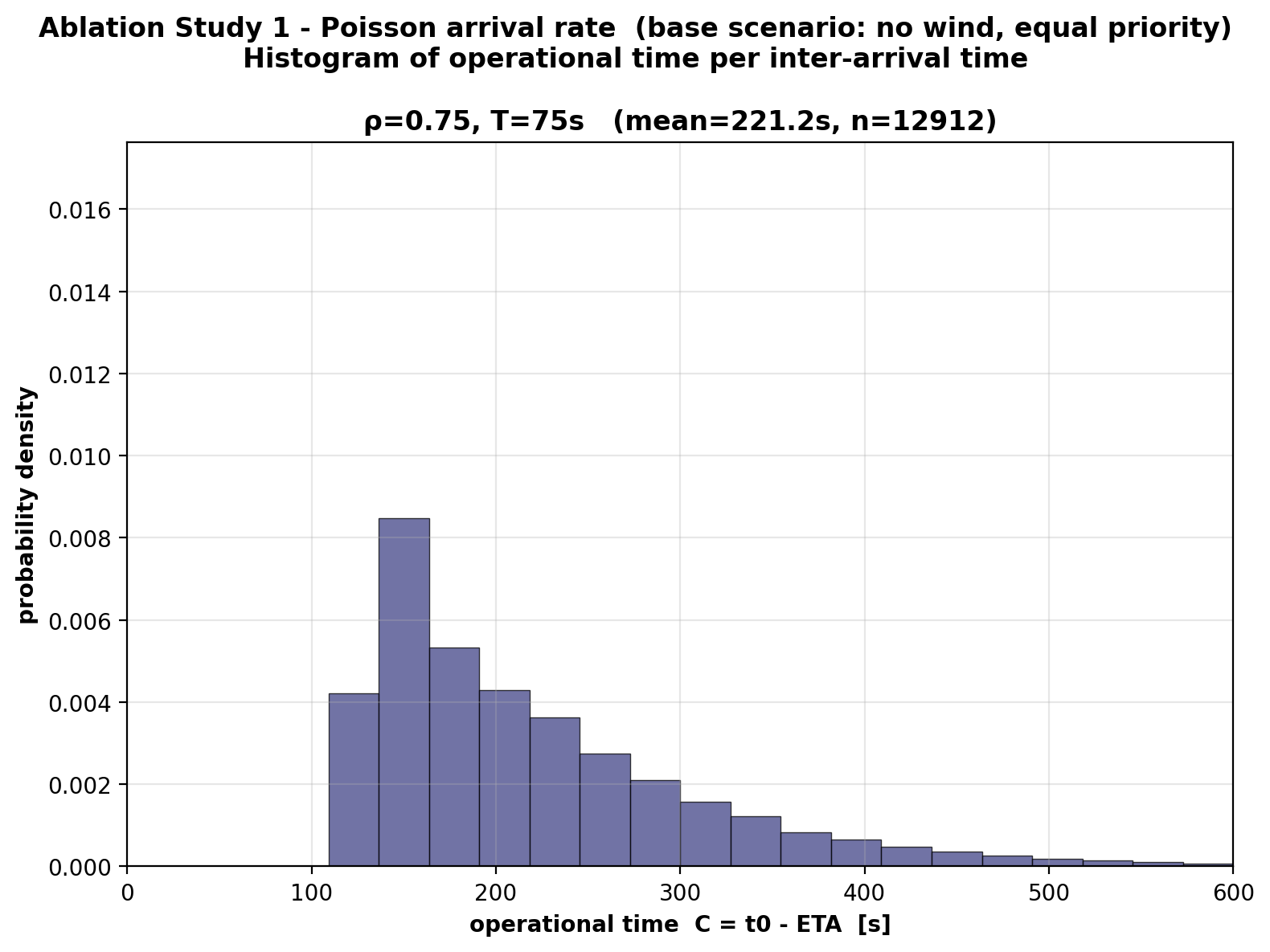}
    \label{fig:histohram2}
}
\subfloat[]{%
    \includegraphics[width=0.32\linewidth]{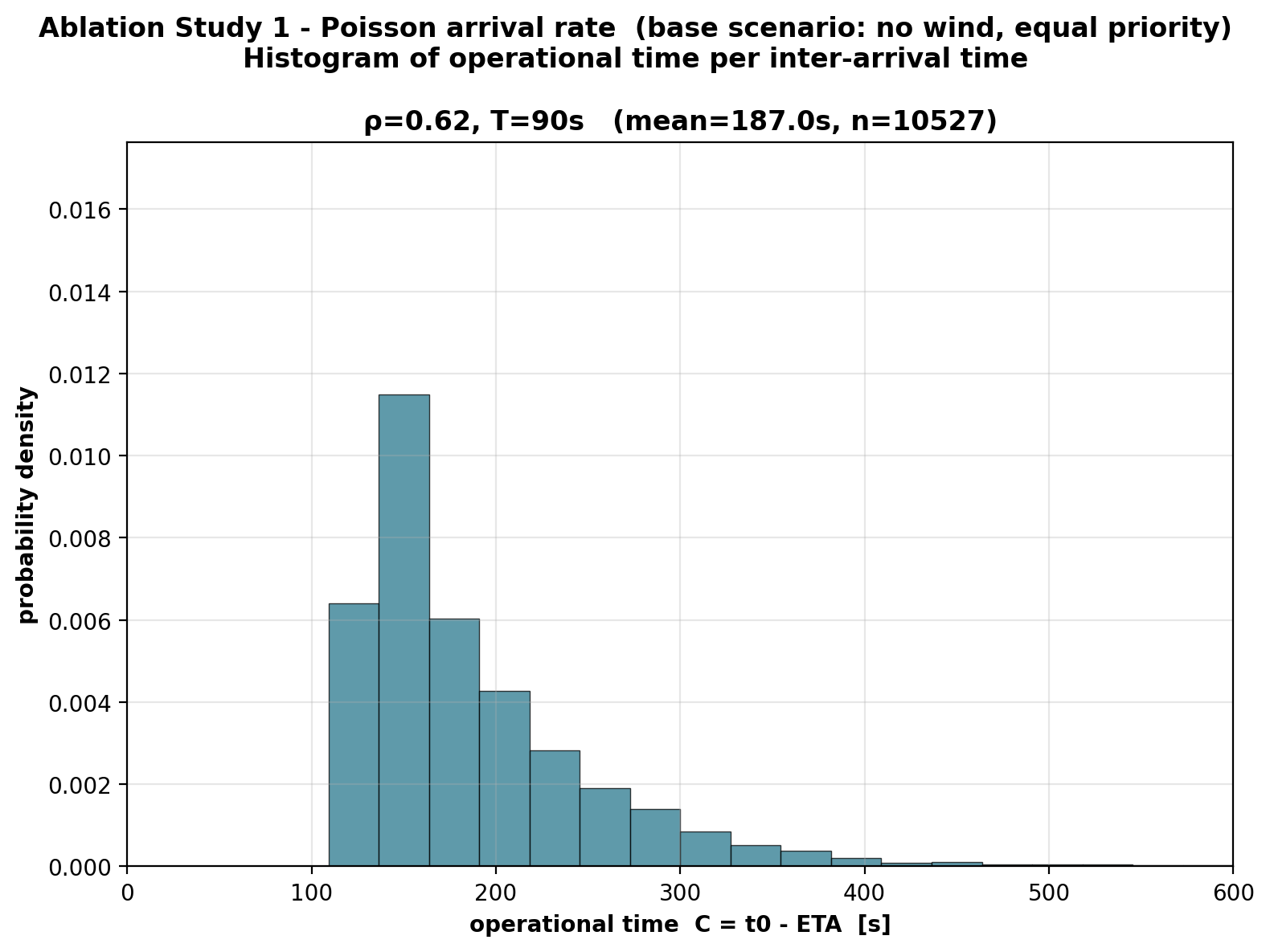}
    \label{fig:histohram3}
}\\
\subfloat[]{%
    \includegraphics[width=0.32\linewidth]{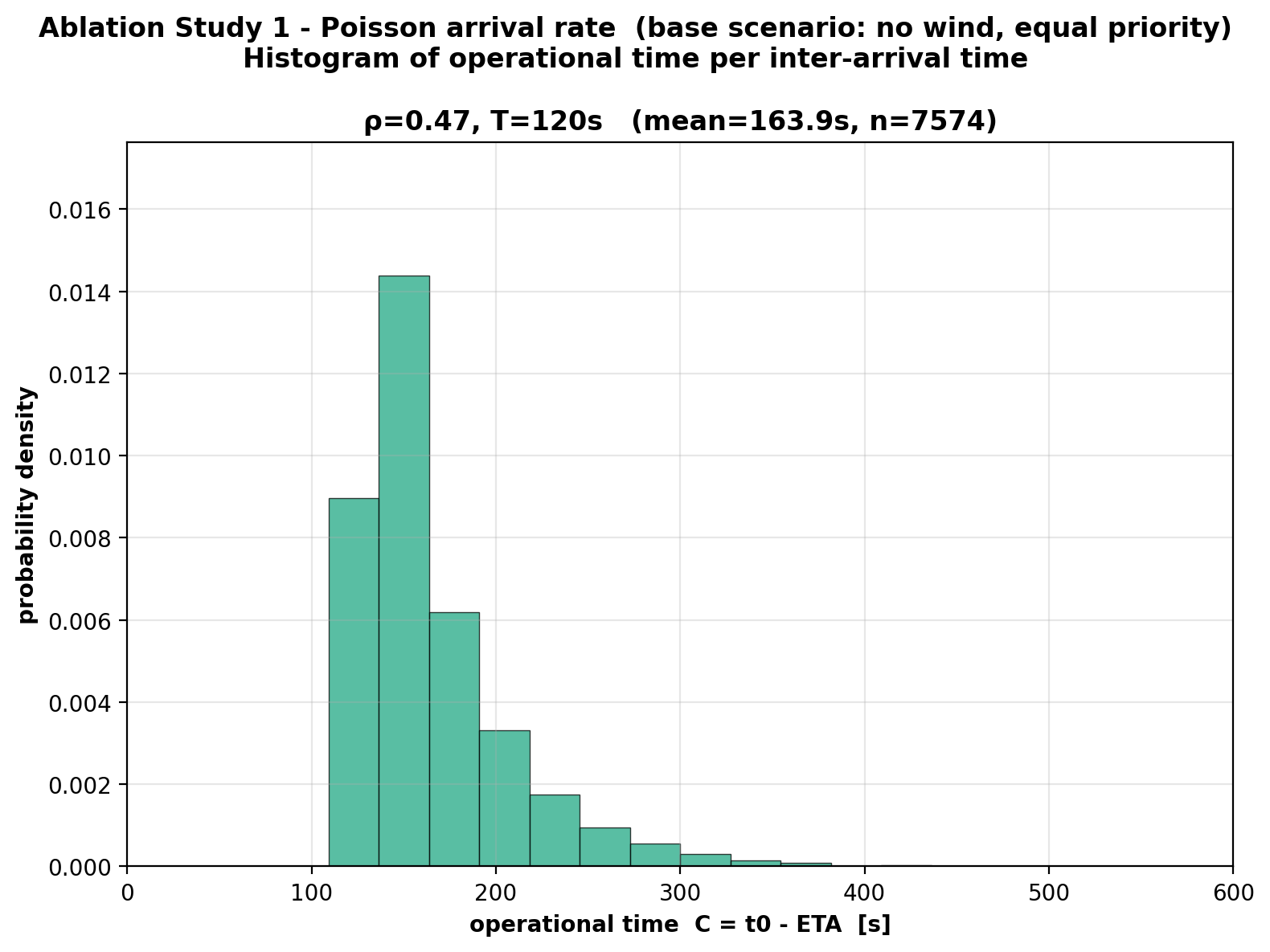}
    \label{fig:histohram4}
}
\subfloat[]{%
    \includegraphics[width=0.32\linewidth]{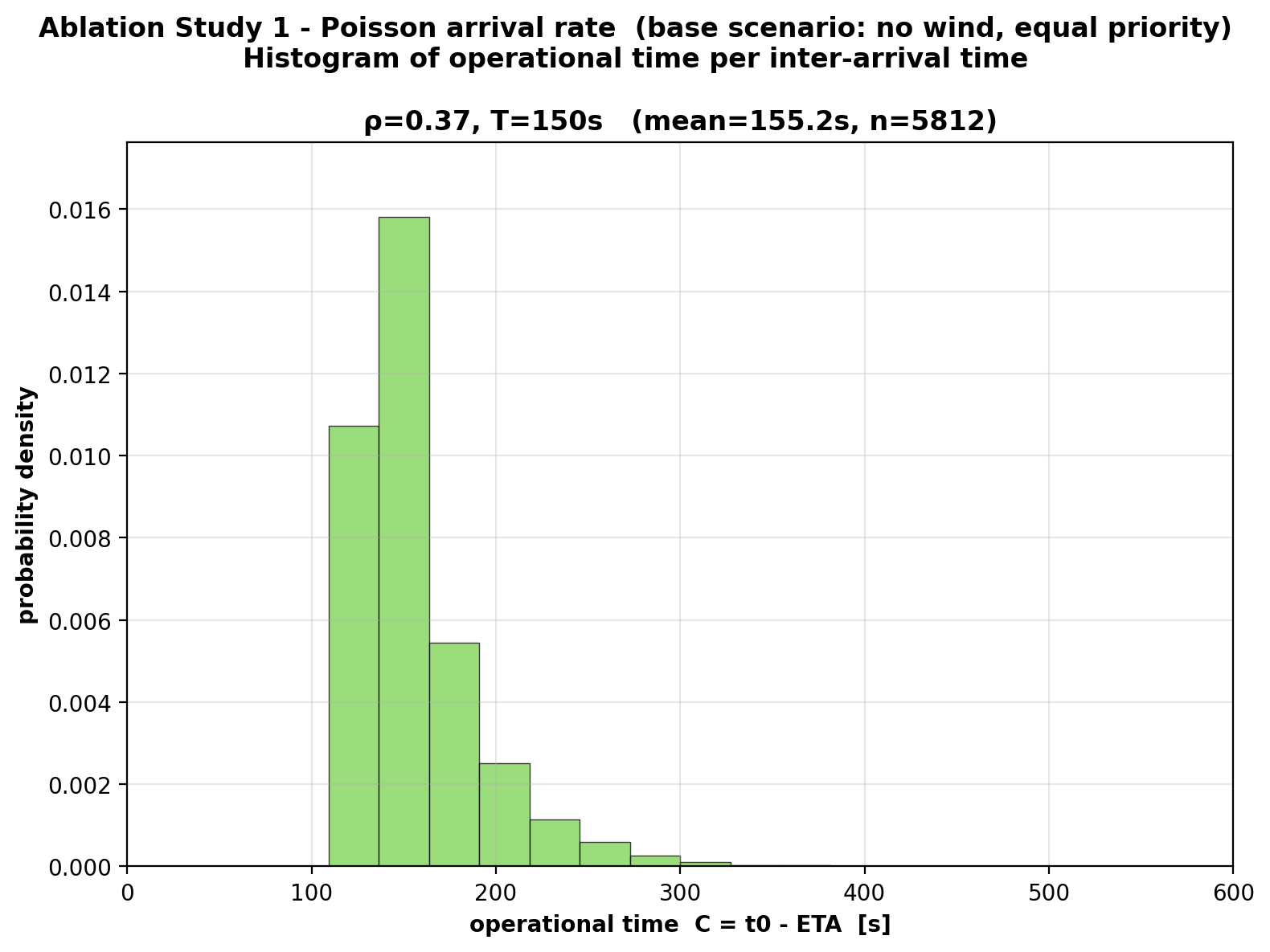}
    \label{fig:histohram5}
}
\subfloat[]{%
    \includegraphics[width=0.32\linewidth]{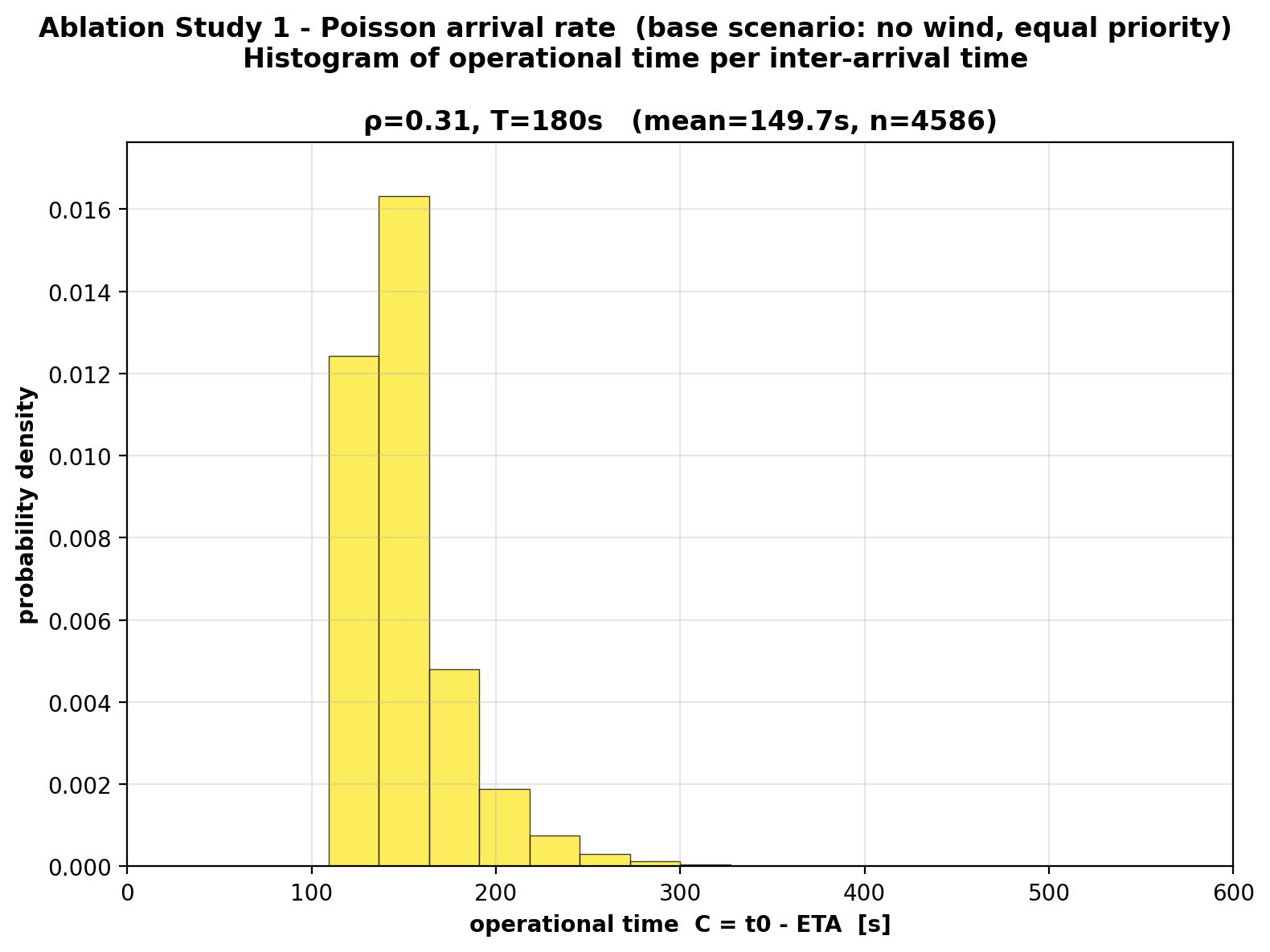}
    \label{fig:histohram6}
}
\caption{Distribution of operational times for various arrival load}
\label{fig:Histogram_arrival_rate}
\end{figure*}

\begin{figure}[t]
    \centering
    \includegraphics[width=0.9\linewidth]{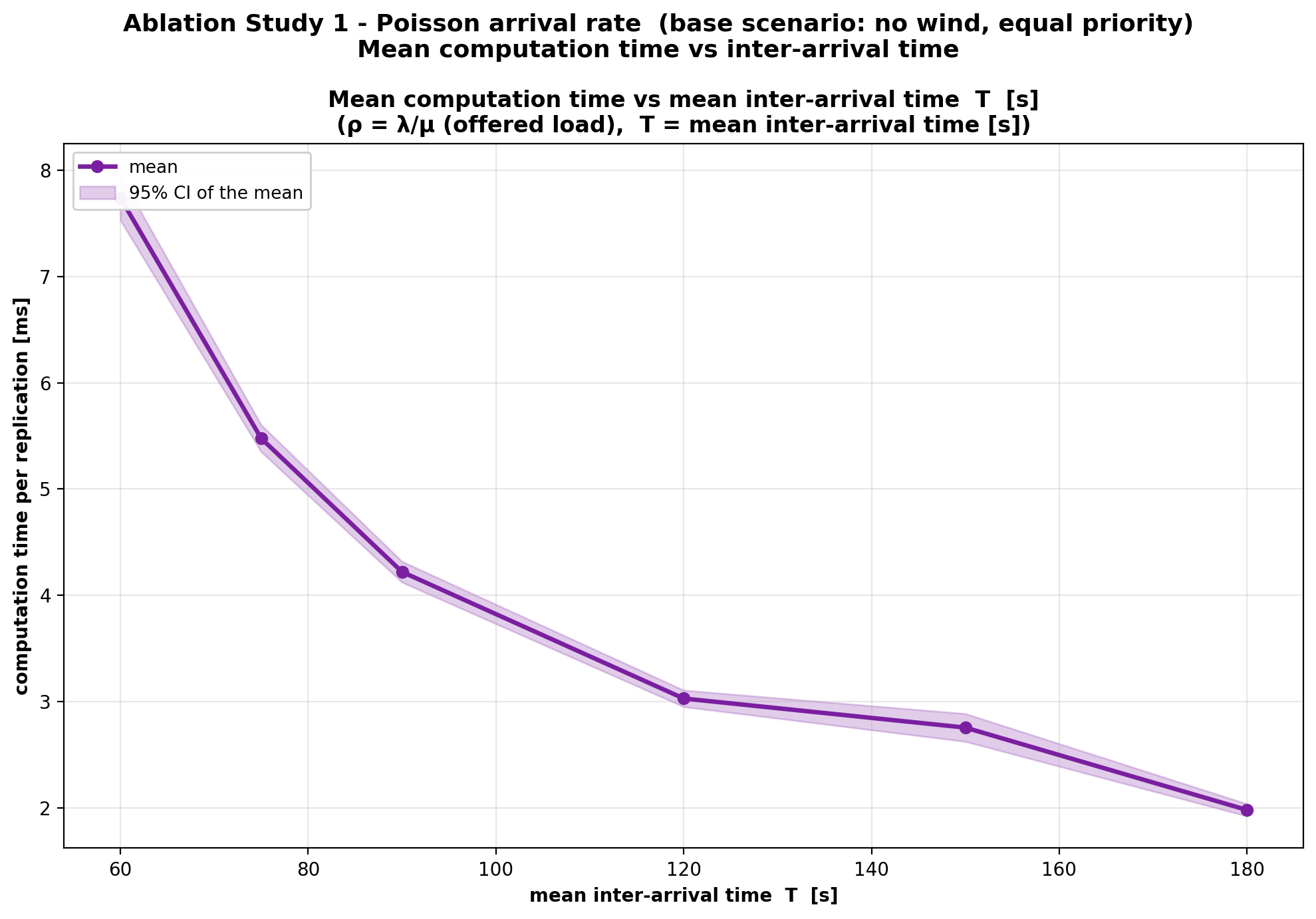}
    \caption{Computational time required for proposed method }
    \label{fig:computational_time_poison}
\end{figure}

The effect of traffic demand is investigated by varying the mean UAS inter-arrival time $T$, where a smaller value of $T$ corresponds to a higher arrival rate. Figure~\ref{fig:operational_time_poison} shows that the normalized operational time increases with the arrival rate. Under high traffic demand, multiple UAS simultaneously compete for the available ring slots, connecting edges, and landing pad, resulting in additional holding and scheduling delays. The corresponding operational-time distributions are presented in Fig.~\ref{fig:Histogram_arrival_rate}. These distributions are positively skewed and qualitatively resemble a gamma distribution. At mean inter-arrival time $T=60$~s, the histogram exhibits a wide spread and a pronounced right tail, indicating substantial variability and the occurrence of comparatively long operational times. As $T$ increases, fewer UAS are simultaneously present in the terminal airspace, reducing resource conflicts and producing more concentrated distributions with shorter right tails. A similar trend is observed in the computational performance shown in Fig.~\ref{fig:computational_time_poison}. At shorter inter-arrival times, the planning method must resolve more overlapping operations and tightly constrained resource assignments, resulting in greater computational effort. Increasing $T$ reduces these interactions and relaxes the temporal constraints, thereby decreasing both the UAS operational time and the computation time.

\subsubsection{Effect of the Number of Rings}

The effect of the terminal-airspace configuration is evaluated by comparing systems with two and three rings. For the three-ring configuration, the distributions of normalized operational time are examined at mean inter-arrival times of $T=60$~s and $T=75$~s. In both cases, the distributions are shifted toward higher operational times compared with the two-ring configuration. This increase is primarily caused by the additional distance and traversal time associated with the third ring, which raises the minimum time required for a UAS to reach the landing pad. However, the three-ring configuration produces a more concentrated distribution with a shorter right tail. The additional ring provides greater spatial and temporal flexibility for assigning holding slots and resolving conflicts before UAS reach the inner layers, thereby reducing the variability and excessive delays caused by resource competition. Although the maximum operational time remains higher for the three-ring system, this is largely attributable to its greater baseline traversal time rather than a higher frequency of severe congestion delays. Thus, adding a ring introduces a trade-off, it increases the nominal operational time while improving traffic organization and reducing the dispersion of conflict-induced delays. 

\begin{figure*}[htbp]
\centering
\subfloat[]{%
    \includegraphics[width=0.32\linewidth]{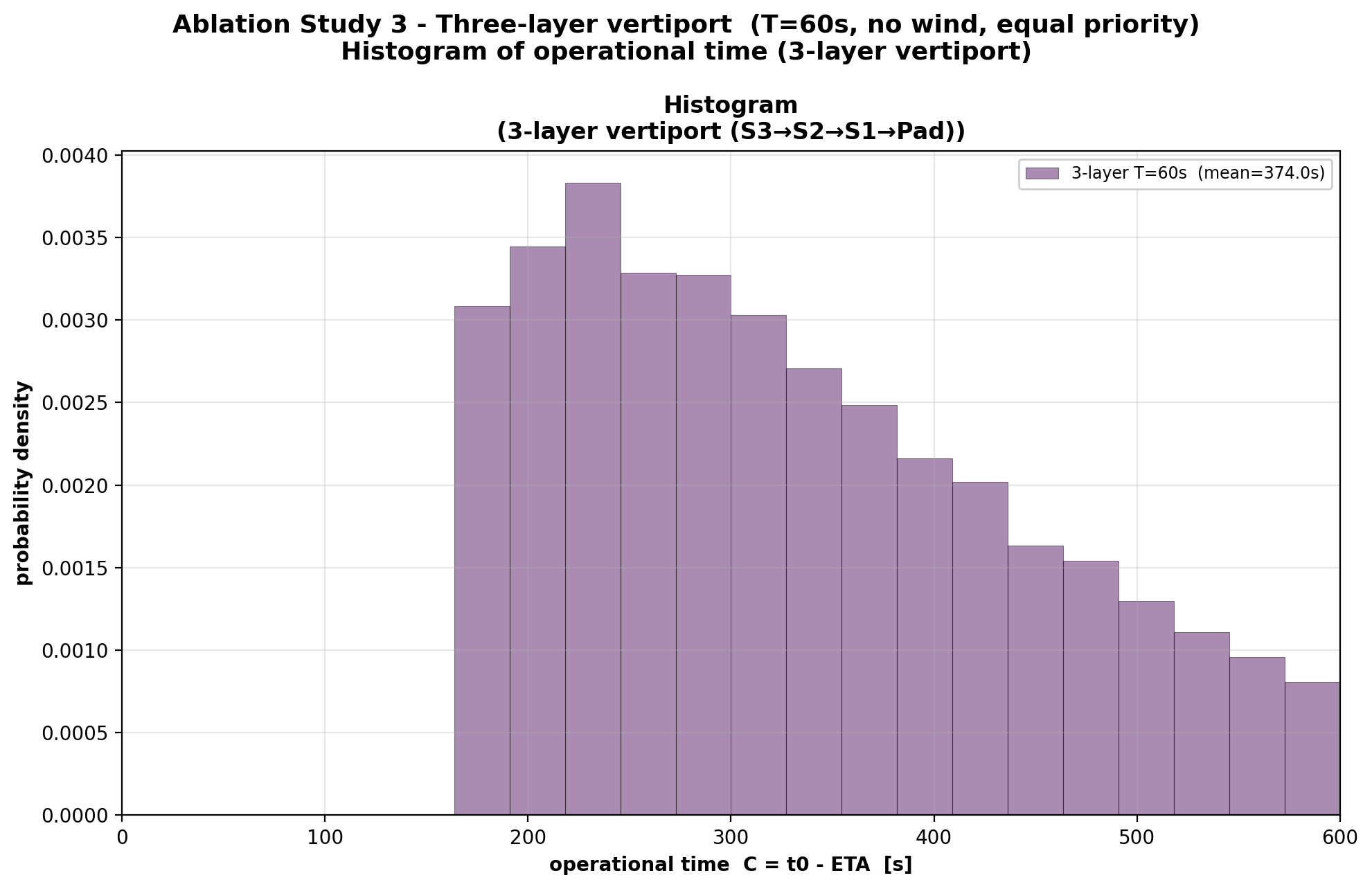}
    \label{fig:3layer_poison_60}
}
\subfloat[]{%
    \includegraphics[width=0.32\linewidth]{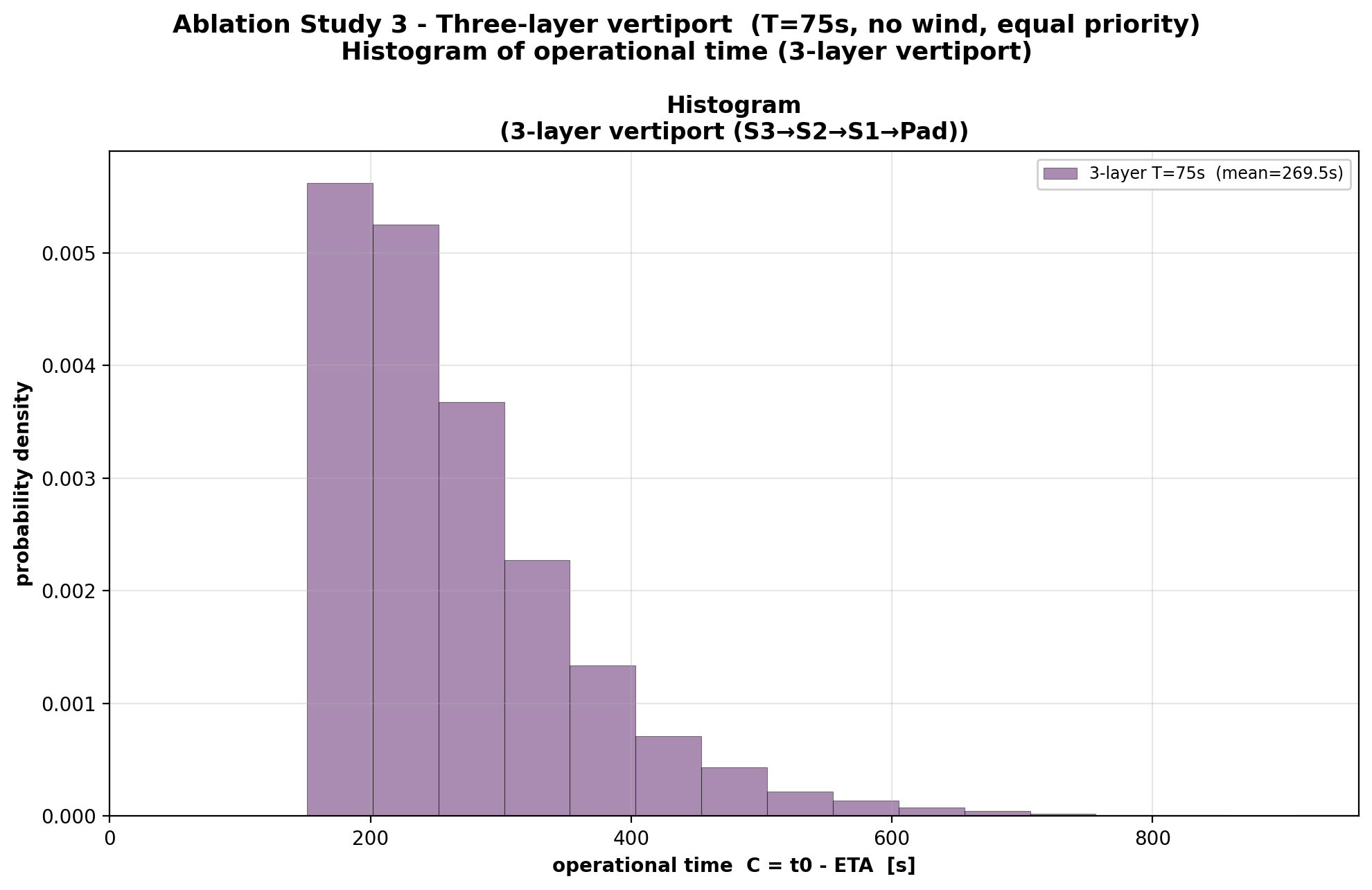}
    \label{fig:3layer_poison_75}
}
\caption{Distribution of operational times for 3-Layer }
\label{fig:3layer_arrival_rate}
\end{figure*}

\subsubsection{Effect of the Spatial Arrival Distribution}
\begin{figure*}[!t]
\centering

\begin{minipage}{0.96\textwidth}
    \centering

    \subfloat[]{%
        \includegraphics[width=0.32\linewidth]
        {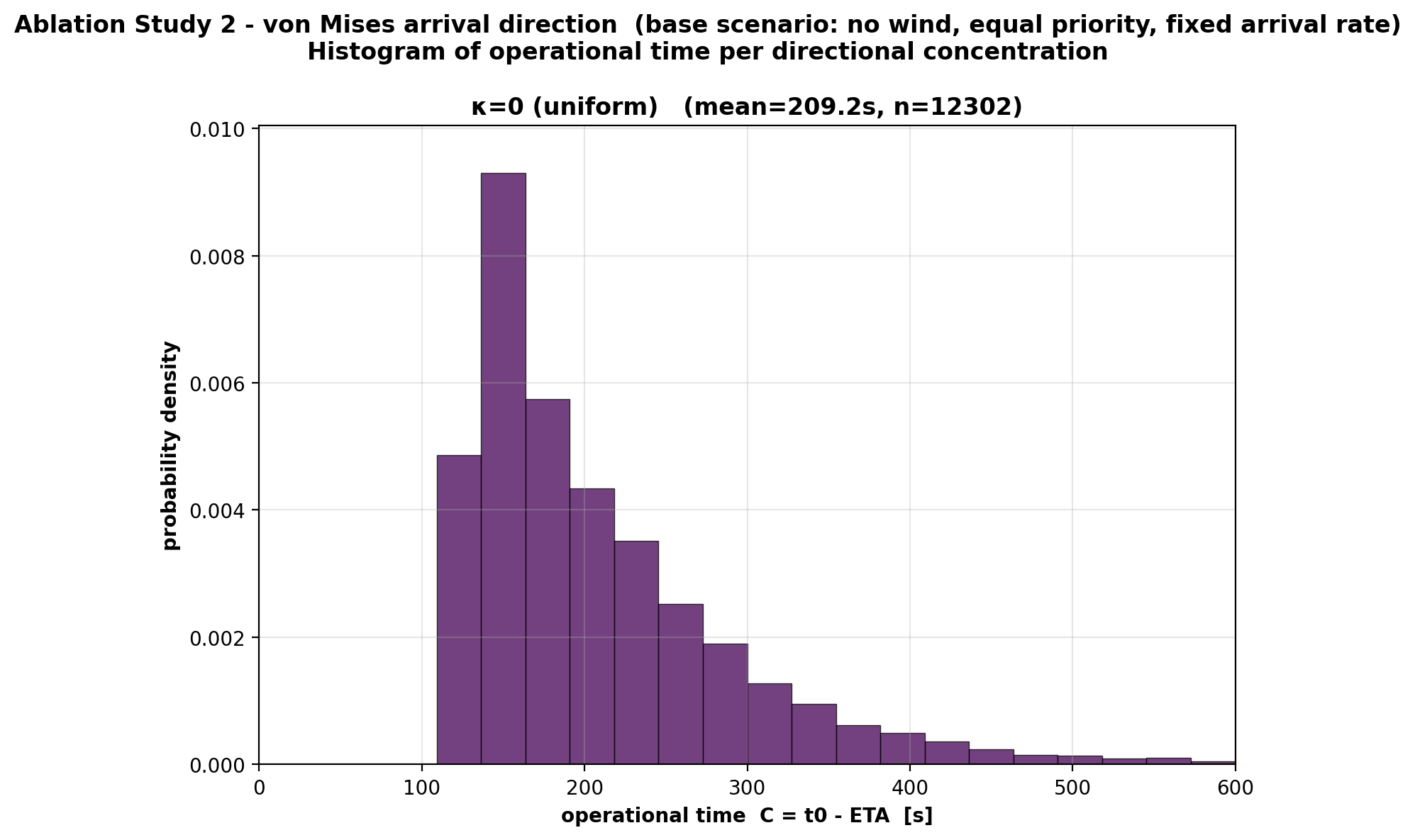}%
        \label{fig:van_histogram_1}%
    }
    \hfill
    \subfloat[]{%
        \includegraphics[width=0.32\linewidth]
        {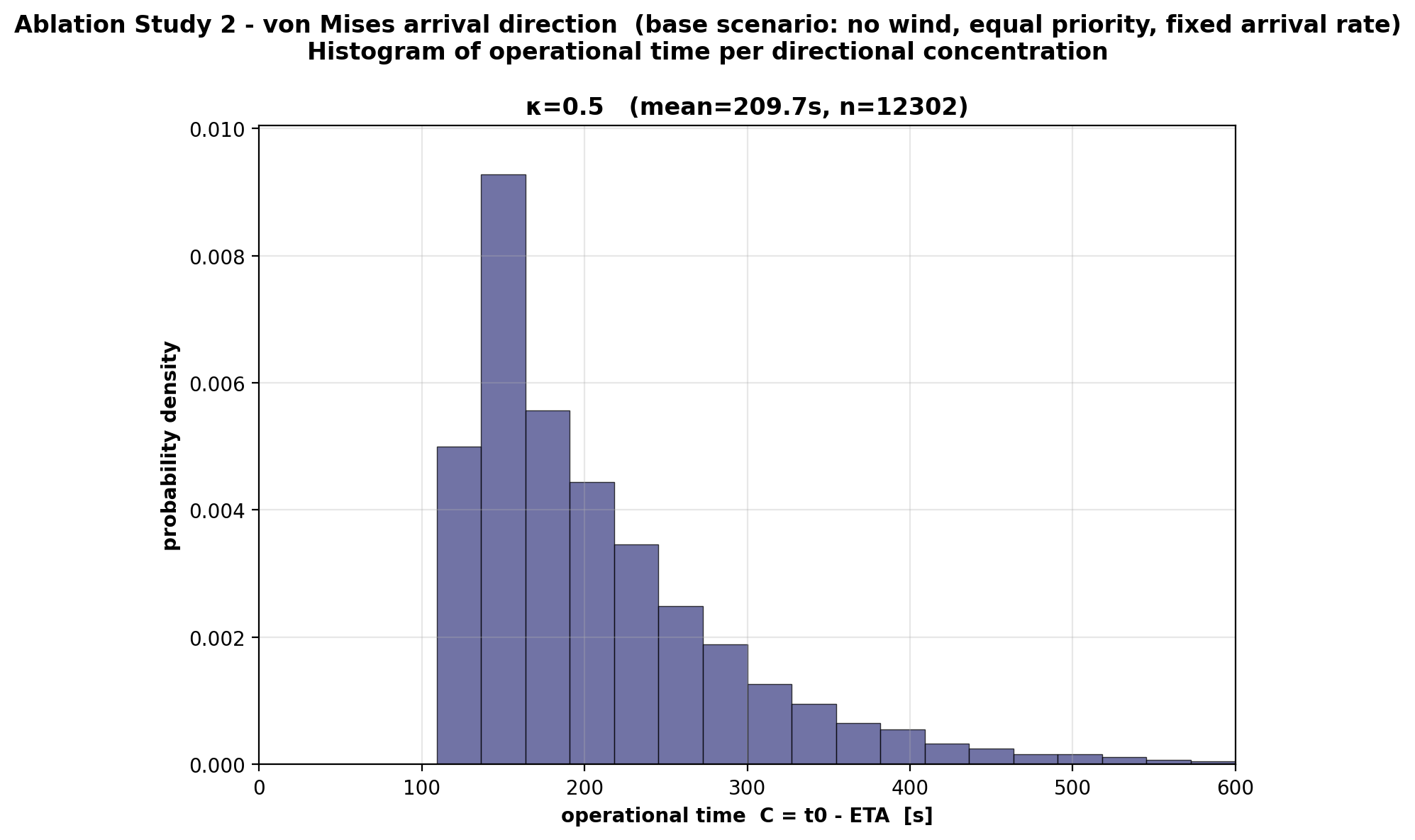}%
        \label{fig:van_histogram_2}%
    }
    \hfill
    \subfloat[]{%
        \includegraphics[width=0.32\linewidth]
        {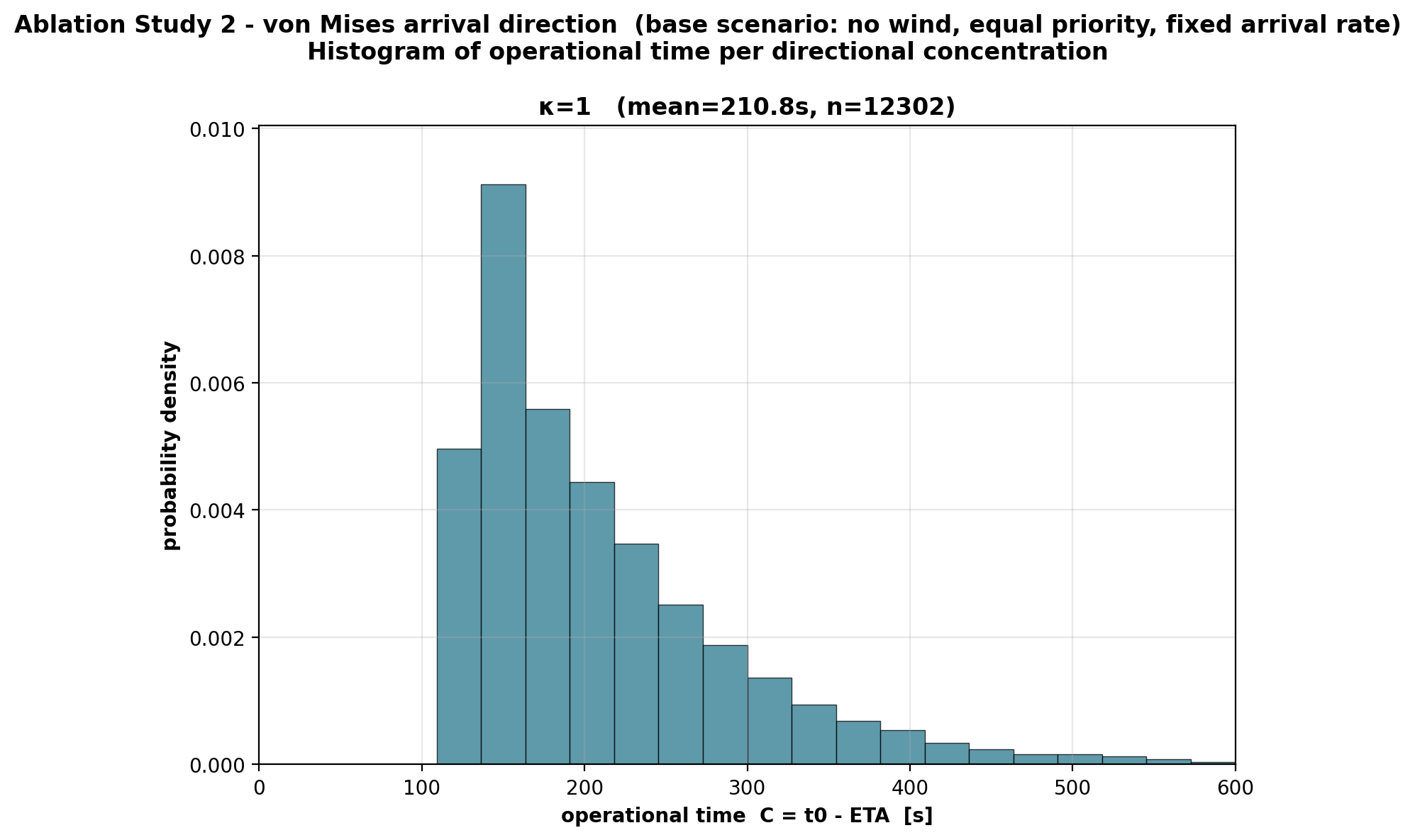}%
        \label{fig:van_histogram_3}%
    }

    \medskip

    \subfloat[]{%
        \includegraphics[width=0.32\linewidth]
        {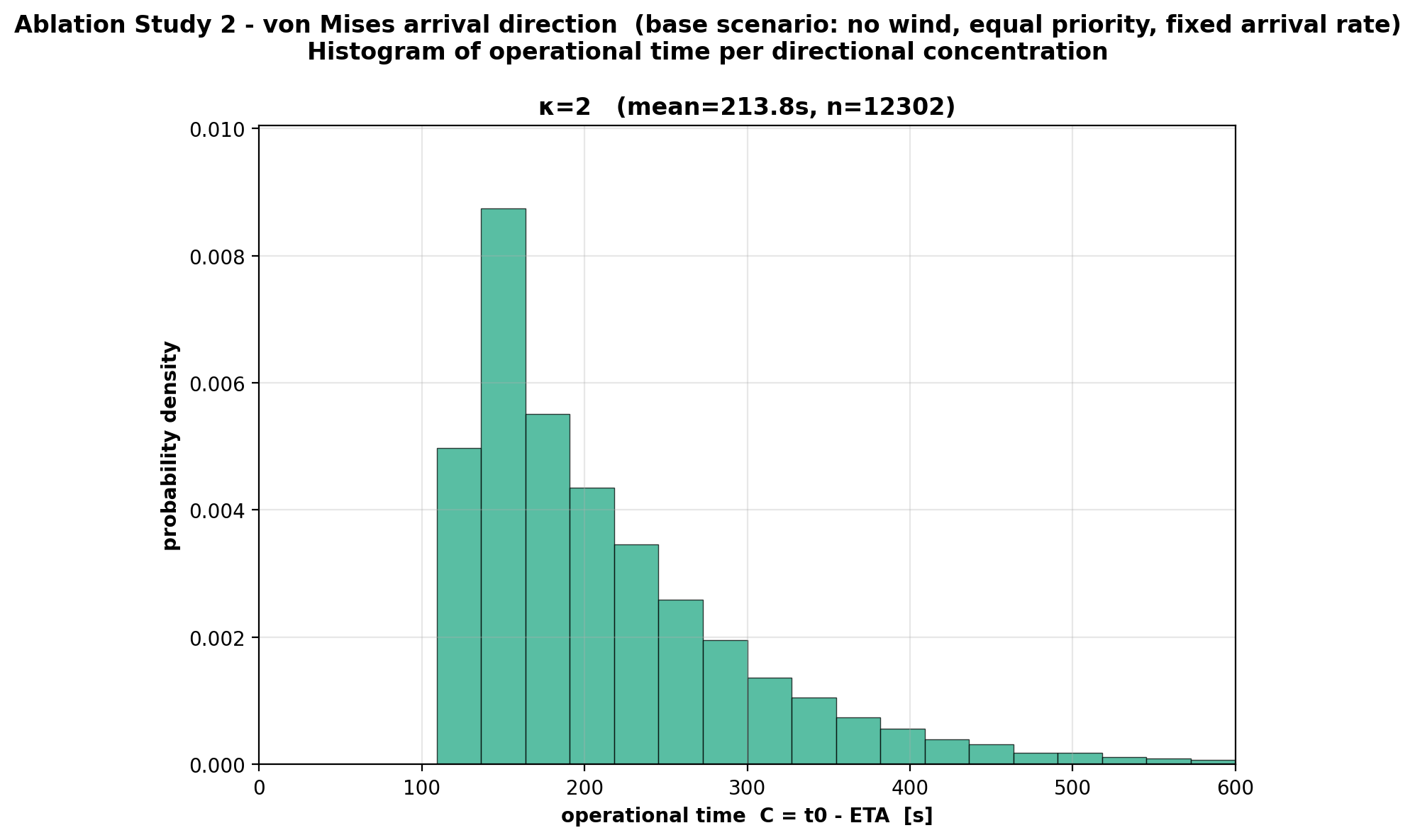}%
        \label{fig:van_histogram_4}%
    }
    \hfill
    \subfloat[]{%
        \includegraphics[width=0.32\linewidth]
        {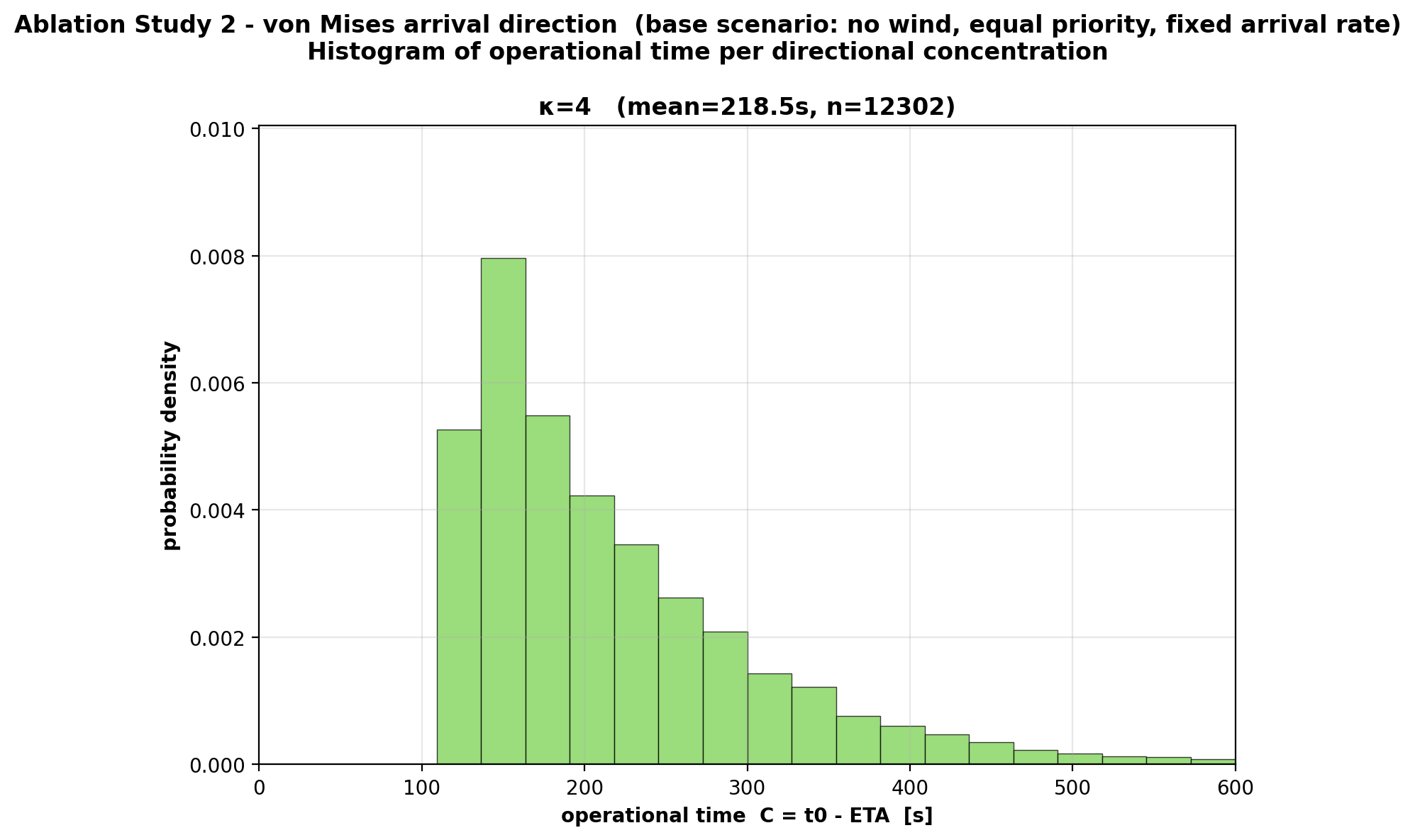}%
        \label{fig:van_histogram_5}%
    }
    \hfill
    \subfloat[]{%
        \includegraphics[width=0.32\linewidth]
        {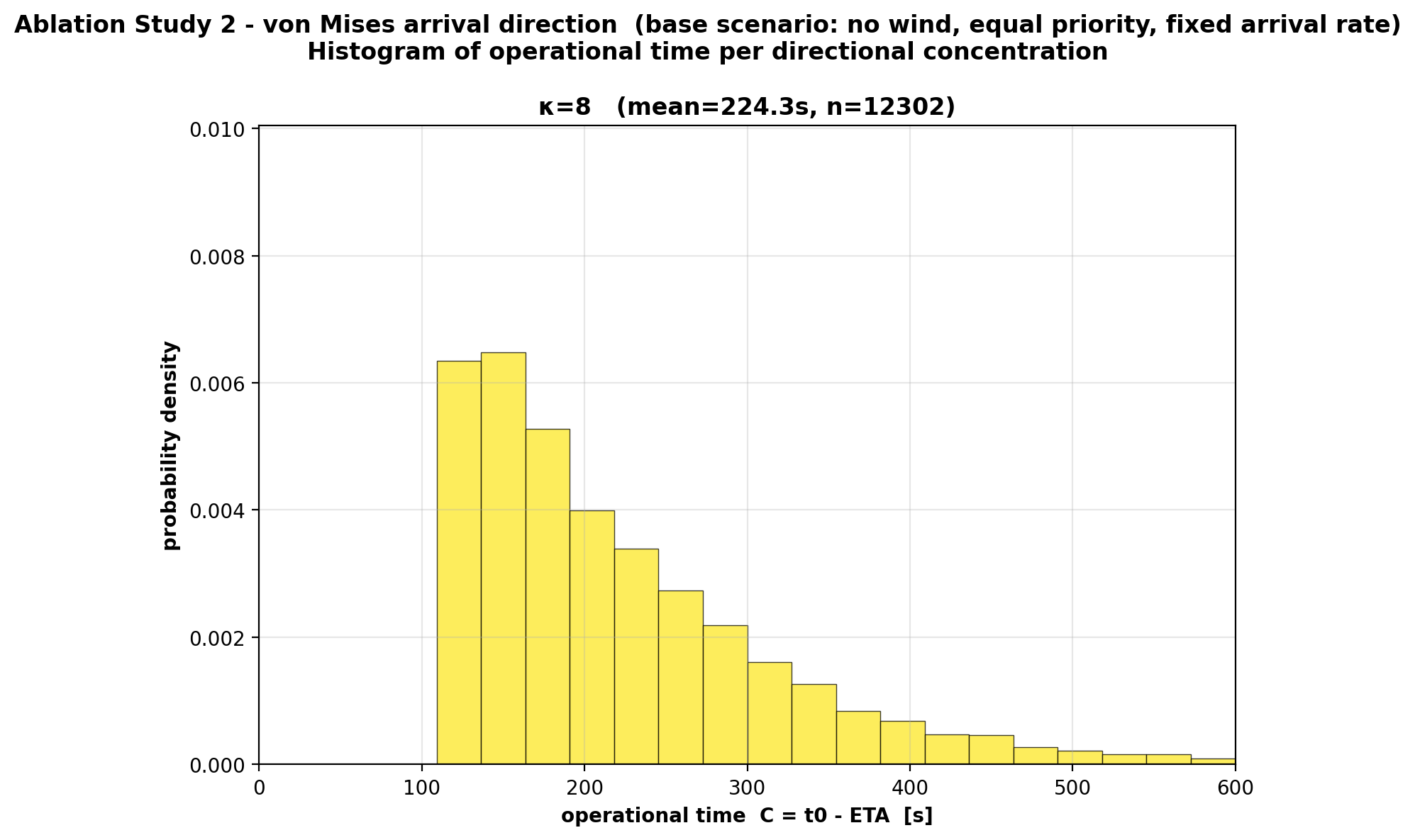}%
        \label{fig:van_histogram_6}%
    }

    \captionof{figure}{%
        Distributions of normalized UAS operational time for
        different spatial arrival distributions.%
    }
    \label{fig:van_histogram}
\end{minipage}

\vspace{4mm}

\begin{minipage}[t]{0.475\textwidth}
    \centering
       \includegraphics[width=\linewidth]{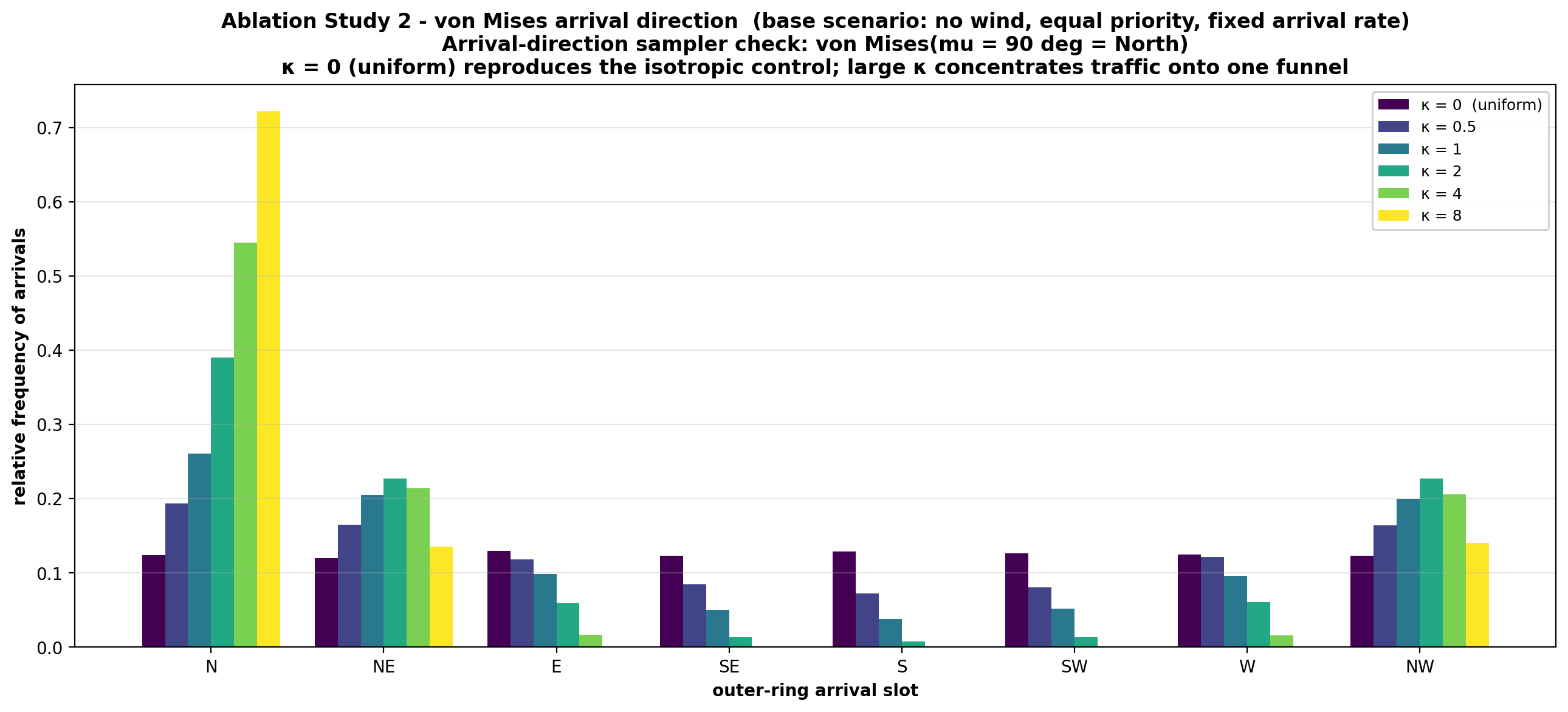}
    \captionof{figure}{Directional distributions of UAS arrivals for different values of $\kappa$.}
    \label{fig:von_distribution}    
\end{minipage}
\hfill
\begin{minipage}[t]{0.475\textwidth}
    \centering
    \includegraphics[width=\linewidth]
    {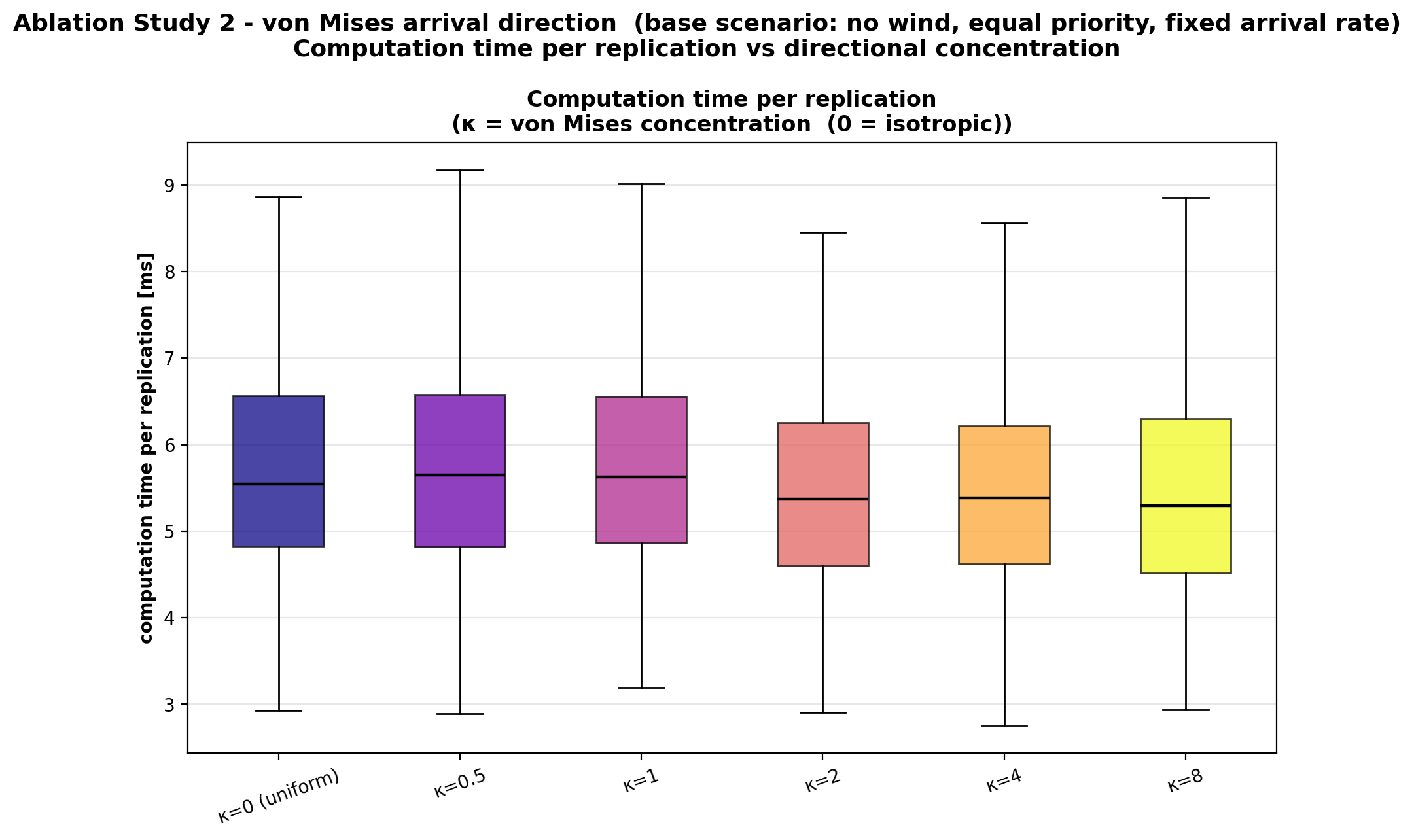}

    \captionof{figure}{%
        Computational time for different spatial arrival distributions.%
    }
    \label{fig:van_computation_time}
\end{minipage}

\end{figure*}

The effect of spatially nonuniform traffic is evaluated by generating UAS arrival directions using a von Mises distribution. The concentration parameter $\kappa$ controls the directional bias: $\kappa=0$ represents uniformly distributed arrivals, whereas larger values concentrate arrivals around a common direction, as shown in Fig.~\ref{fig:von_distribution}. Figure~\ref{fig:van_histogram} shows that the normalized operational-time distributions retain similar shapes, ranges, and central tendencies across the tested values of $\kappa$. Although directional clustering causes more UAS to enter through neighboring outer-ring slots, the scheduler assigns the available slots and transition paths without significantly increasing holding or scheduling delays. The computational times reported in Fig.~\ref{fig:van_computation_time} also remain comparable across the considered distributions. 

Overall, these results demonstrate that the framework is robust to the tested spatial arrival patterns and that its performance is affected more strongly by the number of UAS simultaneously competing for terminal resources than by their arrival directions.

\section{Conclusion}
This work presented a real-time, priority-aware framework for UAS arrival planning in multi-layer UTM terminal airspace. The optimization model jointly determines slot assignments, traversal paths, arrival sequences, and entry and exit times while ensuring conflict-free use of ring slots, connecting edges, and the landing pad. Using a rolling-horizon strategy, the framework continuously updates arrival plans to incorporate new traffic, reactively accommodate wind-induced disturbances, and prioritize emergency landings without disrupting completed operations. The computational results demonstrate that feasible schedules can be generated and updated within operational time limits, supporting real-time decision-making under dynamic conditions. Emergency UAS are granted timely access to constrained resources, while affected trajectories are safely rescheduled following wind disturbances. Overall, the proposed framework provides a computationally practical, responsive, and safety-oriented solution for priority-aware UAS arrival management in dynamic multi-layer UTM environments.


\bibliography{reference.bib}

\begin{thebibliography}{26}
\newcommand{\enquote}[1]{``#1''}
\providecommand{\natexlab}[1]{#1}
\providecommand{\url}[1]{\texttt{#1}}
\providecommand{\urlprefix}{URL }
\expandafter\ifx\csname urlstyle\endcsname\relax
  \providecommand{\doi}[1]{\discretionary{}{}{}https://doi.org/#1}\else
  \providecommand{\doi}[1]{\discretionary{}{}{}\urlstyle{rm}\url{https://doi.org/#1}}\fi

\bibitem[{Acharya et~al.(2026)Acharya, Raza, Vasiloff, Wang, Sun, and Song}]{acharya2026demand}
Acharya, K., Raza, W., Vasiloff, K., Wang, Z., Sun, L., and Song, H.~H., \enquote{Demand Modeling for Advanced Air Mobility: Challenges, Opportunities, and Future Directions,} \emph{IEEE Transactions on Intelligent Transportation Systems}, 2026.

\bibitem[{Wei et~al.(2021)Wei, Nilsson, and Coogan}]{wei2021scheduling}
Wei, Q., Nilsson, G., and Coogan, S., \enquote{Scheduling of urban air mobility services with limited landing capacity and uncertain travel times,} \emph{2021 American Control Conference (ACC)}, IEEE, 2021, pp. 1681--1686.

\bibitem[{Rigas et~al.(2024)Rigas, Kolios, and Ellinas}]{rigas2024scheduling}
Rigas, E.~S., Kolios, P., and Ellinas, G., \enquote{Scheduling aerial vehicles in large scale urban air mobility schemes with vehicle relocation,} \emph{IEEE Transactions on Intelligent Vehicles}, Vol.~9, No.~9, 2024, pp. 5665--5679.

\bibitem[{Tang and Xu(2023)}]{tang2023incorporating}
Tang, Y., and Xu, Y., \enquote{Incorporating optimization in strategic conflict resolution for UAS traffic management,} \emph{IEEE Transactions on Intelligent Transportation Systems}, Vol.~24, No.~11, 2023, pp. 12393--12405.

\bibitem[{Meng et~al.(2025)Meng, Wu, Wen, Zhou, and Tian}]{meng2025optimization}
Meng, L., Wu, M., Wen, X., Zhou, Z., and Tian, Q., \enquote{Optimization of Flight Scheduling in Urban Air Mobility Considering Spatiotemporal Uncertainties,} \emph{Aerospace}, Vol.~12, No.~5, 2025, p. 413.

\bibitem[{Tan et~al.(2019)Tan, Wang, Ong, and Low}]{tan2019evolutionary}
Tan, Q., Wang, Z., Ong, Y.-S., and Low, K.~H., \enquote{Evolutionary optimization-based mission planning for UAS traffic management (UTM),} \emph{2019 International Conference on Unmanned Aircraft Systems (ICUAS)}, IEEE, 2019, pp. 952--958.

\bibitem[{Pradeep and Wei(2019)}]{pradeep2019energy}
Pradeep, P., and Wei, P., \enquote{Energy-efficient arrival with rta constraint for multirotor evtol in urban air mobility,} \emph{Journal of Aerospace Information Systems}, Vol.~16, No.~7, 2019, pp. 263--277.

\bibitem[{Wang et~al.(2026)Wang, Peng, Hua, Luo, Xu, Hu, Cheng, and Zhang}]{wang2026spatiotemporally}
Wang, X., Peng, Z., Hua, P., Luo, R., Xu, H., Hu, J., Cheng, H., and Zhang, B., \enquote{Spatiotemporally Decoupled Hierarchical Reinforcement Learning for Cooperative Multi-UAV Task Allocation and Path Planning,} \emph{2026 International Annual Conference on Complex Systems and Intelligent Science (CSIS-IAC)}, IEEE, 2026, pp. 345--350.

\bibitem[{Pinto~Neto et~al.(2025)Pinto~Neto, Baum, Almeida~Jr, Camargo~Jr, and Cugnasca}]{pinto2025towards}
Pinto~Neto, E.~C., Baum, D., Almeida~Jr, J. R.~d., Camargo~Jr, J.~B., and Cugnasca, P.~S., \enquote{Towards Planning Urban Air Mobility (UAM) Landing Trajectories in Emergencies,} \emph{Journal of Intelligent \& Robotic Systems}, Vol. 111, No.~1, 2025, p.~22.

\bibitem[{Altun et~al.(2023)Altun, Xu, Inalhan, and Hardt}]{altun2023comprehensive}
Altun, A.~T., Xu, Y., Inalhan, G., and Hardt, M.~W., \enquote{Comprehensive risk assessment and utilization for contingency management of future AAM system,} \emph{AIAA AVIATION 2023 Forum}, 2023, p. 3687.

\bibitem[{Kim et~al.(2026)Kim, Sharma, Castagno, Neogi, and Atkins}]{kim2026runtime}
Kim, J., Sharma, P., Castagno, J., Neogi, N., and Atkins, E., \enquote{Runtime performance verification of assured contingency landing management for advanced air mobility,} \emph{Journal of Aerospace Information Systems}, Vol.~23, No.~2, 2026, pp. 155--173.

\bibitem[{S{\'a}ez et~al.(2025)S{\'a}ez, Guitart, Viry, Rad, Prats, Delahaye, and Gonzalez}]{saez2025automated}
S{\'a}ez, R., Guitart, A., Viry, B., Rad, I.~O., Prats, X., Delahaye, D., and Gonzalez, P., \enquote{An automated airborne support tool for aircraft emergencies: selection of landing sites and 4D diversion trajectories,} \emph{IEEE Transactions on Intelligent Transportation Systems}, Vol.~26, No.~4, 2025, pp. 5030--5048.

\bibitem[{Jiang et~al.(2026)Jiang, Li, Li, and Zheng}]{jiang2026bi}
Jiang, B., Li, Y., Li, C., and Zheng, Y., \enquote{Bi-Level Optimization Framework for Urban Low-Altitude UAV Delivery Ensuring Target Level of Safety,} \emph{IEEE Transactions on Intelligent Transportation Systems}, 2026.

\bibitem[{Chen et~al.(2024)Chen, Wei, Krois, Block, Cobb, Louis-Ferdinand, and Hyman}]{chen2024integrated}
Chen, S., Wei, P., Krois, P., Block, J., Cobb, P., Louis-Ferdinand, A., and Hyman, J., \enquote{Integrated arrival and departure management for urban air mobility vertiport operations in the new york city airspace,} \emph{2024 Integrated Communications, Navigation and Surveillance Conference (ICNS)}, IEEE, 2024, pp. 1--10.

\bibitem[{Saxena et~al.(2026)Saxena, Kuri, Prabhakar, and Yadav}]{saxena2026vertiport}
Saxena, R.~R., Kuri, J., Prabhakar, T., and Yadav, M., \enquote{Vertiport terminal scheduling and throughput analysis for multiple surface directions,} \emph{The Aeronautical Journal}, Vol. 130, No. 1350, 2026, pp. 2434--2460.

\bibitem[{Baleghi and Malaek(2025)}]{baleghi2025real}
Baleghi, B., and Malaek, S. M.-B., \enquote{Real-Time Intelligent Landing-Management Under Urban Unpredictable Operations,} \emph{IEEE Transactions on Intelligent Transportation Systems}, 2025.

\bibitem[{Zhang et~al.(2025)Zhang, Grzelak, Lindner, Fricke, and A{\ss}mann}]{zhang2025bigraph}
Zhang, T., Grzelak, D., Lindner, M., Fricke, H., and A{\ss}mann, U., \enquote{A Bigraph-Based Digital Twin for Multi-UAV Landing Management,} \emph{Drones}, Vol.~10, No.~1, 2025, p.~12.

\bibitem[{Vascik and Hansman(2020)}]{vascik2020allocation}
Vascik, P.~D., and Hansman, R.~J., \enquote{Allocation of airspace cutouts to enable procedurally separated small aircraft operations in terminal areas,} \emph{AIAA Aviation 2020 forum}, 2020, p. 2905.

\bibitem[{Hu et~al.(2024)Hu, Pang, and Feroskhan}]{hu2024airspace}
Hu, X., Pang, B., and Feroskhan, M., \enquote{Airspace reconfiguration for urban air mobility: a spatial-temporal analysis of airspace availability within aerodrome,} \emph{2024 IEEE 27th International Conference on Intelligent Transportation Systems (ITSC)}, IEEE, 2024, pp. 1296--1301.

\bibitem[{Chang et~al.(2026)Chang, Hu, Yin, Xie, and Wang}]{chang2026vertiport}
Chang, Z., Hu, M., Yin, J., Xie, H., and Wang, S., \enquote{Vertiport arrival airspace design, analysis, and capacity assessment: integrating the point merge system into UAM,} \emph{Aerospace Science and Technology}, 2026, p. 113051.

\bibitem[{Li et~al.(2025)Li, Zhang, Yi, Deng, and Zhou}]{li2025capacity}
Li, J., Zhang, H., Yi, J., Deng, C., and Zhou, J., \enquote{Capacity assessment of vertiports with a scheduling algorithm considering quality of service,} \emph{The Aeronautical Journal}, Vol. 129, No. 1337, 2025, pp. 1791--1809.

\bibitem[{Song et~al.(2021)Song, Yeo, and Moon}]{song2021approach}
Song, K., Yeo, H., and Moon, J.-H., \enquote{Approach control concepts and optimal vertiport airspace design for urban air mobility (UAM) operation,} \emph{International journal of aeronautical and space sciences}, Vol.~22, No.~4, 2021, pp. 982--994.

\bibitem[{Pradeep and Wei(2018)}]{pradeep2018heuristic}
Pradeep, P., and Wei, P., \enquote{Heuristic approach for arrival sequencing and scheduling for eVTOL aircraft in on-demand urban air mobility,} \emph{2018 IEEE/AIAA 37th Digital Avionics Systems Conference (DASC)}, IEEE, 2018, pp. 1--7.

\bibitem[{Kleinbekman et~al.(2018)Kleinbekman, Mitici, and Wei}]{kleinbekman2018evtol}
Kleinbekman, I.~C., Mitici, M.~A., and Wei, P., \enquote{eVTOL arrival sequencing and scheduling for on-demand urban air mobility,} \emph{2018 IEEE/AIAA 37th Digital Avionics Systems Conference (DASC)}, IEEE, 2018, pp. 1--7.

\bibitem[{Kleinbekman et~al.(2020)Kleinbekman, Mitici, and Wei}]{kleinbekman2020rolling}
Kleinbekman, I.~C., Mitici, M., and Wei, P., \enquote{Rolling-horizon electric vertical takeoff and landing arrival scheduling for on-demand urban air mobility,} \emph{Journal of Aerospace Information Systems}, Vol.~17, No.~3, 2020, pp. 150--159.

\bibitem[{Wei et~al.(2025)Wei, Xiao, Zhao, and Yuan}]{wei2025optimization}
Wei, Z., Xiao, X., Zhao, X., and Yuan, J., \enquote{Optimization Methods for Unmanned eVTOL Approach Sequencing Considering Flight Priority and Traffic Flow Imbalance,} \emph{Drones}, Vol.~9, No.~6, 2025, p. 396.

\end{thebibliography}

 \end{document}